\documentclass[a4paper]{article}

\usepackage[widelayout,hyperrefblack]{research18}

\usepackage{threeparttable}
\usepackage{diagbox}
\usepackage{array}
\usepackage{boldline}
\usepackage{bbm}
\usepackage{algorithmicx}
\usepackage{algpseudocode}
\usepackage{subfigure}

\allowdisplaybreaks
\usetikzlibrary{datavisualization} 

\usepackage{pgfplots}
\usepgfplotslibrary{units}
\usetikzlibrary{spy,backgrounds}
\usetikzlibrary{decorations.markings}
\usepackage{pgfplotstable}

\renewcommand{\indicatorfunction}{\mathop{}\!\mathbbm{1}}

\renewcommand{\indicatorfunction}{\mathop{}\!\mathbbm{1}}

\providecommand{\keywords}[1]{\textbf{\textit{Index terms---}} #1}

\makeatletter
\def\thanks#1{\protected@xdef\@thanks{\@thanks
		\protect\footnotetext{#1}}}
\makeatother

\newcommand{\llg}[1]{{\color{black}#1}}

\begin{document}

\title{\llg{Distribution Decomposition and Sum-Capacity \\ Results of Two-User Optical Intensity Multiple Access Channels}}

\author{ Longguang Li, Ru-Han Chen, Jing Zhou
	\thanks{This work is supported by the National Natural Science Foundation of China under Grant No. 62071489 and Shanghai Sailing Program under Grant No. 21YF1411000. }
\thanks{L. Li is with Dept.  Communication and Electronic Engineering, East China Normal University, Shanghai, China  (e-mail: lgli@cee.ecnu.edu.cn).} 
	\thanks{R. H. Chen is with the Sixty-Third Research Institute, National University of Defense Technology, Nanjing, China (e-mail: tx\_rhc22@nudt.edu.cn).} 
\thanks{J. Zhou is with Dept. Computer Science and Engineering, Shaoxing University, Shaoxing, China (email: jzhou@usx.edu.cn).}
}

\maketitle

\begin{abstract}
This paper investigates the sum-capacity of two-user optical intensity multiple access channels
with per-user peak- or/and average-intensity constraints. By leveraging tools \llg{from} \llg{the decomposition} of \llg{certain maxentropic distributions}, we derive several \llg{lower} bounds on the sum-capacity. In the high signal-to-noise ratio (SNR) regime, some bounds asymptotically match or approach the sum-capacity, thus \llg{closing} or reducing the existing gaps \llg{to the} high-SNR asymptotic sum-capacity. At moderate SNR, some bounds are also fairly close to the sum-capacity.


\end{abstract}

\section{Introduction}
\label{sec:introduction}
\llg{  With an ever-growing demand for high data-rate wireless access, current wireless network is facing severe congestion of the radio frequency (RF) spectrum and wireless traffic bottleneck. In light of this challenge, it is also necessary to consider the optical spectrum for future wireless communication. With the abundance of the optical spectrum spanning around $3$PHz, optical wireless communication (OWC) has become a promising solution to the RF spectrum scarcity, and been considered as a potential technique in future 6G~\cite{ghassemlooy2015emerging,Hemani2017}. OWC includes, e.g., free-space optical (FSO) communication and visible light communication (VLC). FSO systems utilize lasers for medium- and long-range communication, which has been used for metropolitan area network (MAN) extension~\cite{khalighi2014survey}, back-haul for cellular systems~\cite{douik2016hybrid}, and inter-satellite links~\cite{chan2003optical}. VLC systems utilize light-emitting diodes (LEDs) for short-range communication, which can be used for indoor access points (also known as ``LiFi'')~\cite{haas2015lifi,ayyash2016coexistence}, and vehicular communication applications~\cite{kim2015experimental}. Furthermore, due to its inherent physical security and electromagnetic compatibility, OWC has also found other applications, such as military operations~\cite{menaka2021vision}, underwater monitoring~\cite{zeng2016survey,saeed2019underwater}, and intra/inter-chip communications~\cite{taubenblatt2011optical}.}

\llg{From the consideration of implementation and complexity,} most of current OWC systems use \llg{the} so-called intensity-modulation and direct-detection (IM-DD) transmission scheme. In such systems, information is carried on the modulated intensity of the emitted light, and receivers use photodetectors to measure \llg{the} incoming
optical intensity. A widely adopted
channel model for \llg{the} IM-DD based OWC system is the Gaussian
optical intensity channel, which captures key properties
including nonnegativity of optical intensity, input-independent
additive Gaussian noise, and practical constraints on
limited peak or/and average optical intensity. Based on this model, there \llg{have} been extensive \llg{studies} on \llg{channel capacity}~\cite{lapidothmoserwigger09_7,mckellips04_1,hranilovickschischang04_1,
chaabanrezkialouini17_1,limoserwangwigger20_1,
moserwangwigger18_3,chen2021MISO, chaaban2020capacity} as well as coding and modulation constructions~\cite{basarpanayirciuysalhaas16_1,
  yesilkayabasarmiramirkhanipanayirciuysalhaas17_1,songcheng13_1} in the literature.

In an indoor environment,  \llg{OWC is often used for serving multiple users (or devices) simultaneously. Consequently, there has been an increasing interest in multiuser OWC networks in recent years. There are studies on the fundamental limits of OWC networks from an information-theoretic perspective. Bounds or high/low signal-to-noise ratio (SNR) asymptotics of capacity regions of several multiuser OWC channels, including
multiple access channels (MACs)~\cite{chaaban2017capacity,zhou19_1},
broadcast channels \cite{chaaban16_1,soltani2018capacity}  and interference channels~\cite{zhang2022capacity}, etc., have been established. There are also many results on the optimization of OWC networks, including resources and power allocation techniques~\cite{jiang2017joint,erouglu2017multi}, transmission scheme designs~\cite{ling2018efficient,eroglu2019slow}, interference mitigation~\cite{li2018optimization,ma2018coordinated}, energy transfer and harvesting in VLC networks~\cite{fakidis2016indoor}, and network security~\cite{cho2018securing,yin2017physical}. A comprehensive survey on the OWC networks is given in~\cite{obeed2019optimizing}.}

 In this paper, we consider a two-user MAC with per-user peak- or/and average-intensity constraints. The existing works most related to ours are~\cite{chaaban2017capacity,zhou19_1}. In~\cite{chaaban2017capacity}, inner and outer bounds on the capacity region of the MAC have been
established \llg{under per-user peak- and average-intensity constraints}. In the high SNR regime, the capacity gap between these bounds \llg{is} within $1.53$ bits. The low-SNR asymptotic capacity region has also been characterized. A major technique in~\cite{chaaban2017capacity} is to use the truncated Gaussian input distributions and treat the \llg{interference} between users as Gaussian noises while satisfying intensity constraints. The authors in~\cite{zhou19_1} consider the MAC with \llg{a per-user peak- or average-intensity constraint}. For the
average-\llg{intensity} constrained case, \llg{inner and outer bounds are derived, and they asymptotically match at high SNR}. For the peak-intensity
constrained case, at high SNR the asymptotic capacity gap between the bounds is bounded to within $0.09$ bits. The key
idea in~\cite{zhou19_1} is utilizing \llg{the} capacity results
of two additive noise channels where the noises obey certain \llg{maxentropic} distributions.

This paper focuses on the sum-capacity of a two-user MAC with a per-user peak- or/and average-intensity constraints. \llg{Instead of using truncated Gaussian input or existing capacity results on additive noise channels,} \llg{here we leverage tools from} the decomposition of distributions. \llg{Although analyzing distributions of the sum of random
variables is a well-studied problem in network information theory, its inverse problem, i.e., the decomposition of distributions of random
variables, is less studied. In fact, distribution decomposition is an important topic in probability theory, where the employed mathematics lies between the theory of probability and the theory of functions of complex variables. The classical book~\cite{linnik1977decomposition} gives a comprehensive survey on important results in this field. However, most existing mathematical results are on the decomposability or the decomposition types of general random variables, with less restrictions on the decomposed random variables, which cannot be directly applied to analyze network channels. By taking into consideration  \llg{the given constraints on the support or expectations (first moments)} of decomposed random variables, we propose new decomposition results of certain random variables following maxentropic distributions.} These results can be used to derive new lower bounds on sum-capacity in the two-user MAC. The main contributions of the \llg{paper} are as follows:
\begin{enumerate}
\item \textit{Decomposition of Maxentropic Distributions:} By using the binary expansion of a uniform random variable, we establish that the uniform random variable can be decomposed as a sum of two independent random variables while satisfying \llg{given} constraints on their support. Similarly, an exponentially or a truncated exponentially distributed random variable can be decomposed as a sum of two independent random \llg{variables} while satisfying \llg{given} constraints on their expectations. We also show that a uniform random variable can be decomposed as a sum of a `contracted' uniform random variable and an equi-spaced discrete uniform random variable, and a truncated exponential distribution can be decomposed as a sum of a `contracted' truncated exponential distribution and an equi-spaced truncated geometric distribution. Furthermore, we show that similar conclusions also hold for a discrete uniformly distributed, a geometrically distributed, or a truncated geometrically distributed random variable. We establish that a geometrically distributed random variable can be decomposed as a sum of two independent random variables while satisfying given expectation constraints. Also, a truncated geometrically distributed random variable can be decomposed as a sum of two truncated geometrically distributed random \llg{variables}.
\item \textit{New Capacity Lower Bounds and High-SNR Asymptotics:} With results \llg{on decomposing continuous distributions} and \llg{the} Entropy Power Inequality (EPI), we establish lower bounds on \llg{the} sum-capacity. In the \llg{peak-intensity constrained} channel, derived bounds asymptotically match the asymptotic sum-capacity at high SNR, thus \llg{closing} the existing sum-capacity gap \llg{in~~\cite{zhou19_1}}. In the \llg{average-intensity constrained} channel, we recover the high-SNR asymptotic capacity by proposing a new decomposition method on the exponential distribution. In the peak- and average-intensity constrained channel, our derived bounds in several cases \llg{close} or reduce the existing asymptotic sum-capacity gap at high SNR. Furthermore, with results on discrete distribution decomposition and a new single-user capacity bound, we also derive several lower bounds\llg{, which we numerically verify} give good approximations \llg{to} the sum-capacity at moderate \llg{or low} SNR.
\end{enumerate}

The rest of the paper is organized as follows. We end the introduction with a few notational conventions. Sec.~\ref{sec:channel-model} mainly introduces the model of two-user optical intensity multiple access channels. \llg{Sec.~\ref{sec:decomprv} presents} results on \llg{the decomposition} of \llg{several} continuous and discrete random \llg{variables, respectively}. Sec.~\ref{sec:caparesults} shows several \llg{lower} bounds of sum-capacity under different intensity constraints. The paper is concluded in Sec.~\ref{sec:conclusion}. 

\textbf{Notation:} \llg{Upper case letters such as $X$ are used to denote scalar
random variables taking values in the reals $\mathbb{R}$, while their realizations are typically written in lower case, e.g., $x$}. The support of $X$ is denoted by $\mathsf{supp} X$, expectation of $X$ by $\mathbb{E}[X]$, and characteristic function (c.f.) of $X$  by $\phi_{X}(\cdot)$. Entropy is typeset as
$\mathsf{H}(\cdot)$, differential entropy as $\mathsf{h}(\cdot)$, and mutual
information as $\mathsf{I}(\cdot;\cdot)$. For a real number $a$, we use $\lfloor a \rfloor$ and $\lceil a \rceil$ to denote the floor and ceiling of $a$, respectively. \llg{For two positive integers $a$ and $b$, we use $a | b$ to denote $a$ divides $b$}. We denote Dirac delta function by $\delta (\cdot)$. \llg{For a complex number $a$, we denote its conjugate by $\bar{a}$,} and we denote  $i=\sqrt{-1}$. For a positive integer $k$, we denote the index set $\{0,1,2,\ldots,k\}$ by $[k]$. The logarithmic function
$\log(\cdot)$ denotes the natural logarithm.

\section{Channel Model}
\label{sec:channel-model}

\llg{A single-user discrete-time optical intensity channel impaired by Gaussian noise is given by}
\begin{IEEEeqnarray}{rCl}
Y =X+Z,
\label{eq:singlechannelmodel}
\end{IEEEeqnarray}
where $X$ denotes the channel input, and  $Z$ denotes Gaussian noise with variance $\sigma^2$, i.e., 
\begin{IEEEeqnarray}{c}
Z\sim \mathcal{N}(0,\sigma^2).
\label{gaussiannoise}
\end{IEEEeqnarray}
Since $X$ is proportional to \llg{the} optical intensity, its support must satisfy
\begin{IEEEeqnarray}{c}
\mathsf{supp }X \subset \mathbb{R}^+.
\end{IEEEeqnarray}
Considering the limited dynamic range of LED devices and the requirement \llg{for} illumination quality or energy consumption, the input is usually subject to a peak-intensity constraint:
\begin{IEEEeqnarray}{c}
\textnormal{Pr}( X \leq 1 ) = 1,
\end{IEEEeqnarray}
\llg{or/and an average-intensity constraint:}
\begin{IEEEeqnarray}{c}
\mathbb{E} [X]  \leq \alpha, 
\end{IEEEeqnarray}
where the constant $\alpha \in \left(0,\frac{1}{2}\right]$, denotes the average intensity constraint.\footnote{\llg{When $\alpha \in (1/2,1]$, the channel capacity is equal to that of the channel with average intensity constraint being $1/2$.}} 

In this paper, we mainly consider an optical intensity MAC that has two transmitters and one receiver, and the channel output is given by
\begin{IEEEeqnarray}{rCl}
Y =X_1+X_2+Z,
\label{eq:channelmodel}
\end{IEEEeqnarray}
where $X_j, \,\,j=1,2$, denotes the channel input from User $j$, and $Z$ is distributed according to~\eqref{gaussiannoise}.
Similar \llg{to} \llg{the} single-user channel, the support of $X_j$ must satisfy
\begin{IEEEeqnarray}{c}
\mathsf{supp }X_j \subset \mathbb{R}^+, \,\, j=1,2.
\end{IEEEeqnarray}
We assume the input of the MAC is subject to a per-user peak-intensity constraint:
\begin{IEEEeqnarray}{c}
\textnormal{Pr}( X_j \leq \mathsf{A}_j ) = 1, \,\, j=1,2,
\end{IEEEeqnarray}
or/and subject to a per-user average-intensity constraint:
\begin{IEEEeqnarray}{c}
\mathbb{E} [X_j]  \leq \mathsf{E}_j, \,\, j=1,2,
\end{IEEEeqnarray}
where constants $\mathsf{A}_j$ and $\mathsf{E}_j$ \llg{denote} the peak and average intensity constraint of User $j$, respectively. 

Without loss of generality, \llg{in the peak-intensity constrained case} or the peak- and average-intensity constrained case, we assume $\mathsf{A}_1+\mathsf{A}_2=1$, and \llg{in the average} constrained case, we assume $\mathsf{E}_1+\mathsf{E}_2=1$.

In the peak- and average-intensity constrained MAC, we denote the \llg{average-to-peak-intensity} ratio of each user by
\begin{IEEEeqnarray}{rCl}
\alpha_j = \frac{\mathsf{E}_j}{\mathsf{A}_j}, \quad j=1,2,
\end{IEEEeqnarray}
where $\alpha_j \in \left(0,\frac{1}{2}\right]$, and \llg{we also denote} a weighted  \llg{average-to-peak-intensity ratio by
\begin{IEEEeqnarray}{rCl}
\alpha_{\text{w}} = \mathsf{A}_1\alpha_1+\mathsf{A}_2\alpha_2.
\label{eq:alphasum}
\end{IEEEeqnarray}
It is direct to see $\alpha_{\text{w}} = \mathsf{E}_1+ \mathsf{E}_1$.}
\begin{remark}
For a general model with unnormalized constraints, e.g., 
\begin{IEEEeqnarray}{rCl}
\tilde{Y}=h_1\tilde{X}_1+h_2\tilde{X}_2+\tilde{Z}, 
\end{IEEEeqnarray}
with peak-intensity constraint $(\tilde{\mathsf{A}}_1,\tilde{\mathsf{A}}_2)$, average-intensity constraint $(\tilde{\mathsf{E}}_1,\tilde{\mathsf{E}}_2)$, and Gaussian noise variance $\tilde{\sigma}^2$, its sum-capacity can be directly shown to be the same \llg{as} the normalized model~\eqref{eq:channelmodel}, with peak-\llg{intensity} constraint $\left(\frac{h_1\tilde{\mathsf{A}}_1}{h_1\tilde{\mathsf{A}}_1+h_2\tilde{\mathsf{A}}_2},\frac{h_2\tilde{\mathsf{A}}_2}{h_1\tilde{\mathsf{A}}_1+h_2\tilde{\mathsf{A}}_2} \right)$, average-intensity constraint $\left(\frac{h_1\tilde{\mathsf{E}}_1}{h_1\tilde{\mathsf{A}}_1+h_2\tilde{\mathsf{A}}_2},\frac{h_2\tilde{\mathsf{E}}_2}{h_1\tilde{\mathsf{A}}_1+h_2\tilde{\mathsf{A}}_2} \right)$, and $\sigma^2 = \frac{\tilde{\sigma}^2}{(h_1\tilde{A}_1+h_2\tilde{A}_2)^2}$. Similar conclusions also hold for the peak- or the average-intensity constrained case.  
\end{remark}
\llg{Since information is carried on the intensity of the optical signal, in this paper we adopt the definition of optical signal-to-noise ratio (SNR) \cite{Riedl2001,Kamran2020} as follows, }
\begin{IEEEeqnarray}{c}
\textnormal{SNR} = \frac{1}{\sigma}.
\label{eq:snrdef}
\end{IEEEeqnarray}
\llg{It should be noted that the denominator term in~\eqref{eq:snrdef} is $\sigma$ rather than $\sigma^2$, which is different from the definition in the conventional RF channels.}

 \llg{The capacity region of the channel model~\eqref{eq:channelmodel} is defined as the closure of the set of achievable rate pairs, which we denote by $\mathcal{C}$. Then the sum-capacity is defined as the maximum of the sum of rate pairs in $\mathcal{C}$, i.e.,}
\begin{IEEEeqnarray}{rCl}
\mathsf{C}_{\text{sum}} = \max \,\, \{  R_1+R_2: (R_1,R_2) \in \mathcal{C}  \}.
\end{IEEEeqnarray}
It is shown in~\cite{elgamalkim11_1} that the sum-capacity can be expressed as
\begin{IEEEeqnarray}{rCl}
\mathsf{C}_{\text{sum}} = \max_{p_{X_1}(\cdot)  p_{X_2}(\cdot)}  I(X_1+X_2;Y),
\end{IEEEeqnarray}
where the maximum is \llg{optimized} over the product of all feasible input distributions $p_{X_j}(\cdot)$ of User $j$. We use $\mathsf{C}_{\text{p-sum}}$, $\mathsf{C}_{\text{a-sum}}$, and $\mathsf{C}_{\text{ap-sum}}$ to denote the sum-capacity of the peak-intensity constrained, \llg{the} average-intensity constrained, and \llg{the} peak- and average-\llg{intensity} constrained MAC, respectively.
\section{Decomposition of Random Variables}
In this section, we present some existing and new results on \llg{the decomposition} of continuous and discrete random variables useful in \llg{the} paper. We first present results on the decomposition of some continuous random variables. The considered random variables follow \llg{uniform}, exponential, or truncated exponential distributions. Then we present results on \llg{the decomposition} of some discrete \llg{random} variables which follow uniform, geometric, or truncated geometric distributions.
\label{sec:decomprv}
\subsection{Decomposition of continuous random variables}

It is known that any real number $a \in [0,1]$ can be uniquely written in the form\footnote{\llg{Strictly speaking, the binary expansions of dyadic rationals, which possess finite binary representations, are not unique. There are two representations of each dyadic rational other than $0$. For example, $3/4$ can be written as $\frac{3}{4}=\frac{1}{2}+ \frac{1}{2^2}+\frac{0}{2^3}+\frac{0}{2^4}+\cdots$, or $\frac{3}{4}=\frac{1}{2}+ \frac{0}{2^2}+\frac{1}{2^3}+\frac{1}{2^4}+\cdots$. To ensure the uniqueness, in this paper we adopt the former one, which means we write the binary expansions of dyadic rationals in the form in which all digits from a certain position on are zeros. It should also be noted that the dyadic rationals are the only numbers whose binary expansions are not unique. More details can be found in~\cite[p. $41$]{ko2012complexity}.} }
\begin{IEEEeqnarray}{c}
a = \sum_{j=1}^{\infty}{\epsilon_j(a)} {2^{-j}},
\end{IEEEeqnarray}
where $\epsilon_j(a)$ is $0$ or $1$. We also define the following index set:
\begin{IEEEeqnarray}{c}
 \mathcal{I}_a = \{j|\epsilon_{j}(a)=1,j\in \mathbb{N}^+\}.
\end{IEEEeqnarray}

We first present results on the decomposition of a uniformly distributed random variable.

\subsubsection{Decomposition of a uniform random variable}

\begin{proposition}
\label{lem1}
Consider a random variable $U$ uniformly distributed on $[0,1]$. \llg{For any $a \in (0,1)$,} $U$ can be decomposed as a sum of two independent random variables, i.e., 
\begin{IEEEeqnarray}{rCl}
U = U_1+U_2,
\end{IEEEeqnarray}
\llg{where the support of $U_1$ and $U_2$ are the subsets of  $[0,a]$ and $[0,1-a]$, respectively. Specifically, }
$U_1=\sum_{j \in \mathcal{I}_a}B_j 2^{-j} $, and $U_2=\sum_{j \in {{\comp{\mathcal{I}}_{a}}}}B_j 2^{-j} $, with $(B_j)_{j \in \mathbb{N}^+}$ being  \llg{independent and identically distributed} Bernoulli random variables with $\textnormal{Pr}(B_j=1)=\frac{1}{2}$.
\end{proposition}

\begin{IEEEproof} 
\llg{We start with the random sequence $(B_j)_{j \in \mathbb{N}^+}$ defined above. Note that the c.f. of $B_j2^{-j} $ is
\begin{IEEEeqnarray}{rCl}
\phi_{B_j2^{-j}} = \frac{1+e^{it2^{-j}}}{2}.
\end{IEEEeqnarray}
Since $(B_j)_{j \in \mathbb{N}^+}$ are independent and  identically distributed, then
\begin{IEEEeqnarray}{rCl}
\phi_{\sum_{j=1}^{n}B_j2^{-j}} &=& \prod_{j=1}^{n}  \frac{1+e^{it2^{-j}}}{2} \\
&=& \frac{2^{-n}}{1-e^{it2^{-n}}}\prod_{j=1}^{n}  ({1+e^{it2^{-j}}})({1-e^{it2^{-n}}}) \\
&=& \frac{2^{-n}(e^{it}-1)}{e^{it2^{-n}}-1}.
\end{IEEEeqnarray}
By letting $n\rightarrow \infty$ we obtain 
\begin{IEEEeqnarray}{rCl}
\phi_{\sum_{j=1}^{\infty}B_j2^{-j}} &=& \frac{e^{it}-1}{it} \\
 &=& \phi_{U}(t).
\end{IEEEeqnarray}
Hence,
\begin{IEEEeqnarray}{rCl}
U=\sum_{j=1}^{\infty}{B_j} {2^{-j}}.
\label{eq:1888}
\end{IEEEeqnarray}
Then by the fact $\mathcal{I}_a \cup  {\comp{\mathcal{I}}_{a}}=\mathbb{N}^+$, we have
\begin{IEEEeqnarray}{rCl}
U&=&\sum_{j \in \mathcal{I}_a}{B_j} {2^{-j}} + \sum_{j \in {\comp{\mathcal{I}}_{a}}}{B_j} {2^{-j}} \\
 &=&U_1+U_2.
\end{IEEEeqnarray}
 By the definitions of $U_1$ and $U_2$, it is direct to see the support of them are the subsets of $[0,a]$ and $[0,1-a]$, respectively. The independence of $U_1$ and $U_2$ is directly derived from the independence of $(B_j)_{j \in \mathbb{N}^+}$. The proof is concluded.}
\end{IEEEproof}
  In the following, we present \llg{another decomposition result} \llg{in} some special cases when $a = \frac{1}{k}$ with \llg{the} integer $k \geq 2 $.
\begin{proposition}
\label{prop1}
\llg{Consider} a random variable $U$ uniformly distributed on $[0,1]$. \llg{Given an integer $k\geq 2$,} $U$ can be decomposed as a sum of two independent random variables, i.e., 
\begin{IEEEeqnarray}{rCl}
U = U_1+U_2,
\end{IEEEeqnarray}
 where $U_1$ follows a uniform distribution on $\left[0,\frac{1}{k}\right]$, and $U_2$ follows a $k$-point discrete uniform distribution with $\mathsf{supp}U_{2}= \left\{\frac{j}{k}|j \in [k-1]\right\}$.
\end{proposition}
\begin{IEEEproof}
\llg{ The c.f. of $U_1$, $U_2$ and $U$ are given by 
\begin{IEEEeqnarray}{rCl}
\label{eq:repeat1}
\phi_{U_1}(t)=  \frac{k(e^{\frac{it}{k}}-1)}{it},
\end{IEEEeqnarray}
\begin{IEEEeqnarray}{rCl}
\phi_{U_2}(t) &=& \frac{e^{it}-1}{k(e^{\frac{it}{k}}-1)},
\end{IEEEeqnarray}
}
\llg{
and
\begin{IEEEeqnarray}{rCl}
\phi_{U}(t) &=& \frac{e^{it}-1}{it}.
\end{IEEEeqnarray}
}
\llg{It is direct to see $\phi_U(t)=\phi_{U_1}(t)\phi_{U_2}(t)$. The proof is concluded.}

\end{IEEEproof}
\begin{remark}
\llg{In~\cite{lewis1967factorisation}, the author shows that the distributions of the decomposed pair of a uniform distribution are either both singular, or one is (absolutely) continuous and the other is discrete. This indicates that if $a$ is a dyadic rational in Proposition~\ref{lem1}, then the decomposed $U_1$ is a discrete random variable, and $U_2$ is a continuous random variable. If $a$ is not a dyadic rational, then the decomposed $U_1$ and $U_1$ are both singular random variables. Proposition~\ref{prop1} shows that when $a$ is \llg{the} inverse of a positive integer, another decomposition method exists such that $U_1$ is continuous, and $U_2$ is discrete.}
\end{remark}
\subsubsection{Decomposition of an exponential random variable}
\label{sec:decomposition}
We first present \llg{the} existing result on the decomposition of an exponential random variable in~\cite{verdu101_1}.
\begin{proposition}[{\cite[Thm.~$1$]{verdu101_1}}]
Consider a random variable $U$ exponentially distributed with parameter $1$, i.e.,
\begin{IEEEeqnarray}{c}
p_U(u)=e^{-u}, \quad u\geq 0.
\end{IEEEeqnarray}
\llg{For any $a \in (0,1)$,} $U$ can be decomposed as a sum of two independent random variables, i.e., 
\begin{IEEEeqnarray}{c}
U = U_1+U_2,
\end{IEEEeqnarray}
where $U_1$ follows an exponential distribution with parameter $\frac{1}{a}$, and the distribution of $U_2$ is
\begin{IEEEeqnarray}{rCl}
p_{U_2}(u)=a\delta(u)+ (1-a) e^{-u}, \quad u \geq 0.
\end{IEEEeqnarray}
\end{proposition}

\llg{
Next, we propose another decomposition of \llg{the} exponential random variable based on its binary  expansion. Before presenting our result, we first present two useful lemmas. Since the proof of some lemma is in German, and to make the paper more self-contained, we give the proofs of these two lemmas in the Appendices.

\begin{lemma}[{\cite[p.2]{kakeya1914partial}}, {\cite[p.1]{hornich1941beliebige}}, {\cite[Thm. 1]{guthrie1988topological}}]
\label{lem:1112}
Consider a convergent series $\sum_{j=0}^{\infty}a_j$ satisfying $a_j>a_{j+1}>0,\,\forall j \in \mathbb{N}$. We denote its sum by $l$, and its $n$th remainder term by
\begin{IEEEeqnarray}{rCl}
r_n= \sum_{j=n+1}^{\infty}a_j.
\end{IEEEeqnarray}
We further denote the set of its subsum by 
\begin{IEEEeqnarray}{rCl}
\mathcal{A}= \left\{\sum_{j=0}^{\infty} \epsilon_j a_j \Big{|} \epsilon_j = 0 \textnormal{ or } 1, j \in \mathbb{N} \right\}.
\end{IEEEeqnarray}
If $a_n \leq r_n, \, \forall n \in \mathbb{N}$, then $\mathcal{A}$ is the closed interval $[0,l]$.
\end{lemma}
\begin{IEEEproof}
See Appendix~\ref{app:lem11}.
\end{IEEEproof}
}
\begin{remark}
\llg{The result in~Lemma~\ref{lem:1112} is a remark given in~\cite{kakeya1914partial} without proof, and is in essence proved by Hornich~\cite{hornich1941beliebige}.} It should be noted that the set of subsums of infinite convergent series may be a closed interval, a finite union of closed intervals, homeomorphic to a Cantor set, or a \llg{mix} of these three (also known as `Cantorval'). \llg{The full topological classification of the sets of subsums is given in~\cite{guthrie1988topological},} and interested readers are referred to~\cite{guthrie1988topological} for more details. 
\end{remark}
\begin{lemma}[{\cite[Sec.~$3$]{marsaglia1971random}}]
\label{eq:lemmexpo}
Given $\lambda \in (0,\infty)$, and consider a random variable $U$ exponentially distributed with parameter $\lambda$, i.e.,
\begin{IEEEeqnarray}{c}
p_U(u)={\lambda}e^{-{\lambda}u}, \quad u\geq 0.
\end{IEEEeqnarray}
Then 
\begin{IEEEeqnarray}{rCl}
U = \sum_{j=-\infty}^{\infty}B_j 2^j,
\label{eq:uexp}
\end{IEEEeqnarray}
where $(B_j)_{j \in \mathbb{N}^+}$ are independently distributed Bernoulli random variables and  $\textnormal{Pr}(B_j=1)=\frac{1}{1+e^{\lambda 2^j}}$.
\end{lemma}
\begin{IEEEproof}
\llg{See Appendix~\ref{app:1a}.}
\end{IEEEproof}
Now we \llg{are ready to} present the result on the decomposition of an exponential random variable.
\begin{proposition}
\label{lem2}
Consider a random variable $U$ exponentially distributed with parameter $1$. \llg{For any $a \in (0,1)$,} there exists a set $\mathcal{I} \subset \mathbb{Z}$ such that $U$ can be decomposed as a sum of two independent random variables, i.e., 
\begin{IEEEeqnarray}{c}
U = U_1+U_2,
\end{IEEEeqnarray}
and
\begin{IEEEeqnarray}{c}
 \mathsf{E}[U_1]=a,
\end{IEEEeqnarray} 
where $U_1=\sum_{j \in \mathcal{I}}B_j  2^{-j} $, and $U_2=\sum_{j \in {\comp{\mathcal{I}}}}B_j  2^{-j} $, with $(B_j)_{j \in \mathbb{N}^+}$ being  independently distributed Bernoulli random variables and $\textnormal{Pr}(B_j=1)=\frac{1}{1+e^{ 2^j}}$.
\end{proposition}
\begin{IEEEproof}
Let $\lambda=1$, and take expectation at both sides in~\eqref{eq:uexp} yielding
\begin{IEEEeqnarray}{c}
\label{eq:kkk}
\sum_{j=-\infty}^{\infty}\frac{2^j}{1+e^{2^j}}=1.
\end{IEEEeqnarray}
It is sufficient to show there exists a set $\mathcal{I} \subset \mathbb{Z}$ such that 
\begin{IEEEeqnarray}{c}
\label{eq:subsum}
\sum_{j\in \mathcal{I}}\frac{2^j}{1+e^{2^j}}=a.
\end{IEEEeqnarray}
To prove this, we apply \llg{Lemma~$1$} to show \llg{that} the set of all subsums of the infinite series in~\eqref{eq:kkk} forms the closed interval $[0,1]$. To this end, we first rearrange and merge some terms of \llg{the} sequence in~\eqref{eq:kkk} to form a new infinite series that \llg{satisfies the} required conditions \llg{in Lemma~$1$}. 

Denote the sequence in~\eqref{eq:kkk} by $\left(a_j= \frac{2^j}{1+e^{2^j}}\right)_{j \in \mathbb{Z} }$, and define a new sequence $(b_n)_{n \in \mathbb{N}}$, where 
\begin{IEEEeqnarray}{rCl}
b_n=\begin{cases}
a_0 &\textnormal{if } n=0, \\
a_{-1} &\textnormal{if } n=1, \\
a_{1}  &\textnormal{if } n=2, \\
a_{-2} &\textnormal{if } n=3, \\
a_{2}  &\textnormal{if } n=4, \\
a_{-n+2}+ a_{n-2} &\textnormal{if } n\geq 5.
\end{cases}
\end{IEEEeqnarray}
Then by ~\eqref{eq:kkk} it is direct to see 
\begin{IEEEeqnarray}{c}
\label{eq:kk2}
\sum_{j=0}^{\infty}b_j =1.
\end{IEEEeqnarray}
\llg{The $n$th remainder term} of \llg{the} above series is
\begin{IEEEeqnarray}{c}
\label{eq:tail}
r_n = \sum_{j=n+1}^{\infty}b_j.
\end{IEEEeqnarray}
It is proved in Appendix~C that \llg{the} sequence $(b_n)_{n \in \mathbb{N}}$ is monotonically decreasing, and satisfies
\begin{IEEEeqnarray}{c}
\label{eq:lessrn}
b_n \leq r_n, \,\, \forall n \in \mathbb{N}.
\end{IEEEeqnarray}
 \llg{By Lemma~$1$}, the set of subsums of the new infinite series in~\eqref{eq:kk2} indeed \llg{forms} a closed interval. Since the least upper bound and greatest lower bound of the series are $0$ and $1$, respectively, the formed interval is $[0,1]$. Hence, there must exist an index set $\mathcal{J}\subset \mathbb{N}$ such that 
\begin{IEEEeqnarray}{c}
\label{eq:subsum2}
\sum_{j\in \mathcal{J}}b_j = a.
\end{IEEEeqnarray}
Then by~\eqref{eq:kkk} there exists a set $\mathcal{I}\subset \mathbb{Z}$ satisfying~\eqref{eq:subsum}. The proof is concluded.
\end{IEEEproof}

\subsubsection{Decomposition of a truncated exponentially distributed random variable}

Here we \llg{present the} decomposition of a truncated exponential random variable based on its binary  expansion, which is characterized by the following lemma\llg{, whose proof} follows the similar arguments as in the proof of Lemma~\ref{eq:lemmexpo}.

\begin{lemma}[{\cite[Thm.~$1$]{marsaglia1971random},~\cite{chatterji1963certain}}]
\label{prop:truncatedexp}
Consider a random variable $U$ following a truncated exponential distributed over support $[0,1]$ with parameter $\lambda > 0$, i.e.,
\begin{IEEEeqnarray}{c}
p_U(u)=\frac{\lambda}{1-e^{-\lambda}} e^{-{\lambda}u}, \quad u \in [0,1].
\end{IEEEeqnarray}
Then 
\begin{IEEEeqnarray}{rCl}
U = \sum_{j=1}^{\infty}B_j  2^{-j},
\label{eq:uexp2}
\end{IEEEeqnarray}
where $(B_j)_{j \in \mathbb{N}^+}$ are independently distributed Bernoulli random variables and  $\textnormal{Pr}(B_j=1)=\frac{1}{1+e^{\lambda 2^{-j}}}$.
\end{lemma}

Based on Lemma~\ref{prop:truncatedexp}, we \llg{present} our result on the decomposition of a truncated exponential random variable.
\begin{proposition}
\label{prop:truncated}
Consider a random variable $U$ following a truncated exponential distributed over interval $[0,1]$ with parameter $\lambda > 0$. \llg{For any  $a \in (0,1)$}, $U$ can be decomposed as a sum of two independent random variables, i.e., 
\begin{IEEEeqnarray}{c}
U = U_1+U_2,
\end{IEEEeqnarray}
with
\begin{IEEEeqnarray}{c}
 \mathsf{E}[U_1]\leq \frac{a}{2},
\end{IEEEeqnarray} 
and 
\begin{IEEEeqnarray}{c}
\label{eq:mmmm}
 \mathsf{E}[U_2]\leq \frac{1-a}{2},
\end{IEEEeqnarray} 
where $U_1=\sum_{j \in \mathcal{I}_a}B_j  2^{-j} $, and $U_2=\sum_{j \in {{\comp{\mathcal{I}}_a}}}B_j  2^{-j} $ with $(B_j)_{j \in \mathbb{N}^+}$ being defined as in \llg{Lemma}~\ref{prop:truncatedexp}.
\end{proposition}
\begin{IEEEproof}
By~\eqref{eq:uexp2} and the fact $\mathcal{I}_a \cup  {\comp{\mathcal{I}}_{a}}=\mathbb{N}^+$, we have
\begin{IEEEeqnarray}{rCl}
U&=&\sum_{j \in \mathcal{I}_a}{B_j} {2^{-j}} + \sum_{j \in {\comp{\mathcal{I}}_{a}}}{B_j} {2^{-j}} \\
 &=&U_1+U_2.
\end{IEEEeqnarray}
The independence of $U_1$ and $U_2$ is directly derived from the independence of $B_j$'s \llg{which are the same as in Lemma~\ref{prop:truncatedexp}}.

The expectation of $U_1$ satisfies 
\begin{IEEEeqnarray}{rCl}
 \mathsf{E}[U_1] = \sum_{j \in \mathcal{I}_a} \frac{2^{-j}} {1+e^{\lambda 2^{-j}}} 
\leq \sum_{j \in \mathcal{I}_a}{2^{-j-1}} 
=\frac{a}{2}.
\end{IEEEeqnarray} 
Besides,~\eqref{eq:mmmm} can be proved by following the same argument.
\end{IEEEproof}

\begin{figure}[htbp]
\subfigure[a uniform distribution]{   
\begin{minipage}{12cm}
\centering    
\includegraphics[scale=0.7]{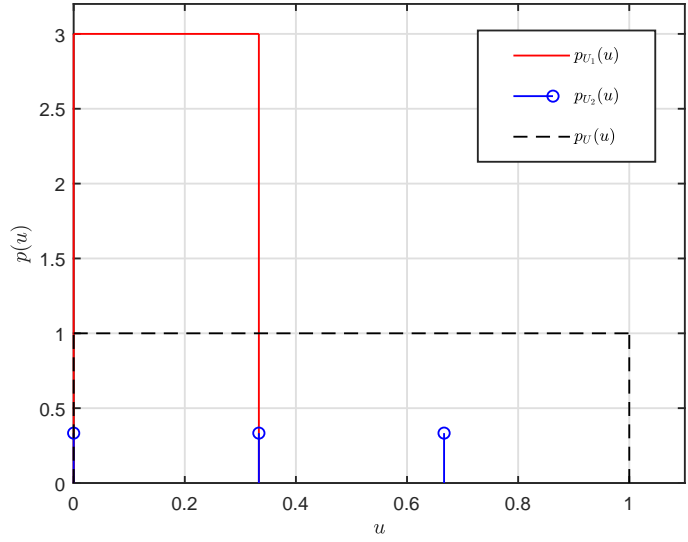}  
\end{minipage}
}
\subfigure[a truncated exponential distribution with $\lambda=1$.]{ 
\begin{minipage}{12cm}
\centering    
\includegraphics[scale=0.7]{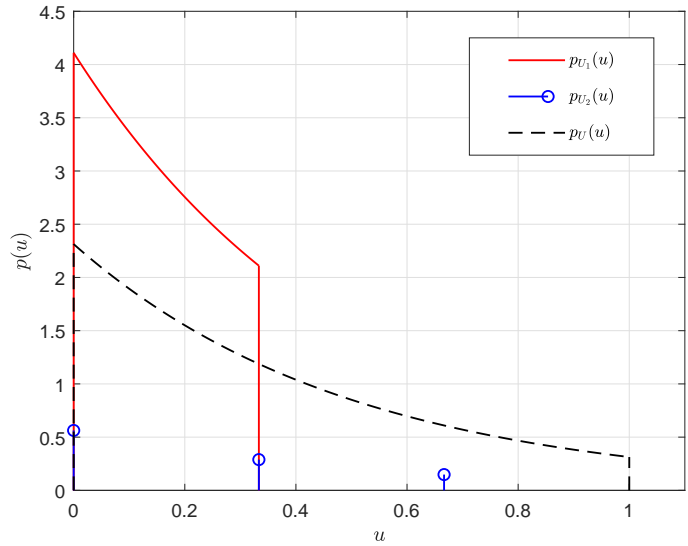}
\end{minipage}
}
\caption{Decomposition of a uniform and a truncated exponential distribution with $k={3}$.}
\label{fig:01}
\end{figure}

For general $a \in (0,1)$, the distributions of $U_1$ and $U_2$ may be singular. In the following, we present \llg{the} result on some special cases when $a = \frac{1}{k}$ with integer $k \geq 2 $.
\begin{proposition}
\label{prop9}
Consider a random variable $U$ following a truncated exponential distributed over interval $[0,1]$ with parameter $\lambda>0$. \llg{Given an integer $k\geq 2$,} $U$ can be decomposed as a sum of two independent random variables, i.e., 
\begin{IEEEeqnarray}{rCl}
U = U_1+U_2,
\end{IEEEeqnarray}
 where $U_1$ follows a truncated exponential distribution over interval $\left[0,\frac{1}{k}\right]$ with parameter $\lambda$, i.e.,
\begin{IEEEeqnarray}{rCl}
p_{U_1}(u)=\frac{\lambda}{1-e^{-\frac{\lambda}{k}}} e^{-\lambda u},\quad\forall u\in \left[0,\frac{1}{k}\right],
\label{eq:u1trunc}
\end{IEEEeqnarray}
and $U_2$ follows a $k$-point truncated geometric distribution with parameter $1-e^{-\frac{\lambda}{k}}$ and $\mathsf{supp}U_{2}= \left\{\frac{j}{k}|j \in [k-1]\right\}$, i.e.,
\begin{IEEEeqnarray}{rCl}
p_{U_2}\left(\frac{j}{k}\right)=\frac{1}{1-e^{-\lambda}}  (1-e^{-\frac{\lambda}{k}}) e^{-\frac{\lambda j}{k}}.
\label{eq:u2trunc}
\end{IEEEeqnarray}
\end{proposition}
\begin{IEEEproof}
\llg{The c.f. of $U_1$, $U_2$, and $U$ are given by 
\begin{IEEEeqnarray}{rCl}
\phi_{U_1}(t)= \frac{\lambda}{1-e^{-\frac{\lambda}{k}}} \frac{e^{\frac{it-\lambda}{k}}-1}{it-\lambda},
\end{IEEEeqnarray}

\begin{IEEEeqnarray}{rCl}
\phi_{U_2}(t)= \frac{1-e^{-\frac{\lambda}{k}}}{1-e^{-\lambda}} \frac{e^{it-\lambda}-1}{e^{\frac{it-\lambda}{k}}-1},
\end{IEEEeqnarray}
and
\begin{IEEEeqnarray}{rCl}
\phi_{U}(t)= \frac{\lambda}{1-e^{-\lambda}} \frac{e^{it-\lambda}-1}{it-\lambda} .
\label{phiutrunc}
\end{IEEEeqnarray}
}
\llg{It is direct to see $\phi_U(t)=\phi_{U_1}(t)\phi_{U_2}(t)$. The proof is concluded.}
\end{IEEEproof}
 Figure~\ref{fig:01} shows an example \llg{of} the \llg{decomposition} of a uniform distribution in Proposition~\ref{prop1} and a truncated exponential distribution with parameter $\lambda=1$ in Proposition~\ref{prop9}. These two distributions can both be decomposed as a sum of a `contracted' version of themselves and an equi-spaced discrete distribution.

\subsection{Decomposition of discrete random variables}
Here we present results on \llg{the} \llg{decomposition} of discrete uniformly distributed, geometrically distributed, and truncated geometrically distributed random variables.
\subsubsection{Decomposition of a discrete uniform random variable}
In~\cite{behrends1999falschen}, Behrends shows \llg{classical} results on \llg{decomposing} a discrete uniform random variable. In the following, we cite some of the results useful in this paper. Since the original paper is in German, the proofs of these results are also given in this paper. We first show a necessary condition on the distributions of the decomposed random variables from a discrete uniform distribution.

\begin{proposition}[{\cite[Lem.~$2.1$]{behrends1999falschen}}]
\label{prop:decompunifdisc}
\llg{Consider} three positive integers $k_1$,~$k_2$, and $k$ with $k=k_1+k_2$. Let $U_{D_1}$ and $U_{D_2}$ be independent random variables whose support are the subsets of $[k_1]$ and $[k_2]$, respectively. If the sum of $U_{D_1}$ and $U_{D_2}$, i.e., $U_D=U_{D_1}+U_{D_2}$, is uniformly distributed over \llg{the} support $[k]$, then $U_{D_1}$ and $U_{D_2}$ must be uniformly distributed over their support.
\end{proposition}

\begin{IEEEproof}
\llg{Since $U_D$ is uniformly distributed over the support $[k]$,} the support of $U_{D_i}$ must contain \llg{the points} $0$ and $k_i$, otherwise it contradicts the fact $p_{U_D}(0)=p_{U_D}(k) = \frac{1}{k+1}$. We now define two polynomials 
\begin{IEEEeqnarray}{rCl}
P(x)=\sum_{j \in [k_1]}a_jx^j, \label{eq:px}
\end{IEEEeqnarray}
and 
\begin{IEEEeqnarray}{rCl}
Q(x)=\sum_{j \in [k_2]}b_jx^j, \label{eq:qx}
\end{IEEEeqnarray}
where $a_j=\frac{p_{U_{D_1}}(j)}{p_{U_{D_1}}(0)}$, $b_j=\frac{p_{U_{D_2}}(j)}{p_{U_{D_2}}(0)}$, \llg{and $x \in \mathbb{C}$}.

\llg{To prove $U_{D_1}$ and $U_{D_2}$ are uniformly distributed over their own support, by the definitions of $a_j$ and $b_j$, it suffices} to prove that $a_j$ in~\eqref{eq:px} and $b_j$ in~\eqref{eq:qx} are either $0$ or $1$. To do this, we first show that $P(x)$ (and $Q(x)$) is palindromic, i.e., $a_j=a_{k_1-j},\,\,\forall j \in [k_1]$.  

 Since $U_D$ is uniformly distributed over \llg{the} set $[k]$, we have 
\begin{IEEEeqnarray}{rCl}
P(x)  Q(x)=\sum_{j \in [k]}x^j. \label{eq:ptimesq}
\end{IEEEeqnarray}
Note that
\begin{IEEEeqnarray}{rCl}
(1-x)  P(x)  Q(x)=1-x^{k+1}. \label{eq:roots}
\end{IEEEeqnarray}
Denote the set of the roots of $P(x)$ by $\Gamma$. Then $\forall \gamma \in \Gamma$, by~\eqref{eq:roots} we have $|\gamma|=1$, hence \llg{its conjugate} $\bar{\gamma}=\frac{1}{\gamma}$. Since all $a_j$'s \llg{are} real numbers, $\bar{\gamma} \in \Gamma$. Note that $P(0)=1$, we express $P(x)$ in terms of its roots, and yield
\begin{IEEEeqnarray}{rCl}
P(x)&=& \prod_{\gamma \in \Gamma}(x-\gamma) \\ 
&=& x^{k_1}\prod_{\gamma \in \Gamma}\left(1-\frac{\gamma}{x}\right) \\
&=& x^{k_1} \prod_{\gamma \in \Gamma}(-\gamma)  \prod_{\gamma \in \Gamma}\left({\frac{1}{x}-\frac{1}{\gamma}}\right) \\
&=& x^{k_1} P(0)  \prod_{\gamma \in \Gamma}\left(\frac{1}{x}-\bar{\gamma}\right) \\
&=& x^{k_1} \prod_{\gamma \in \Gamma}\left(\frac{1}{x}-{\gamma}\right) \\
&=& x^{k_1} P\left(\frac{1}{x}\right). \label{eq:xinverse}
\end{IEEEeqnarray}
Comparing the coefficients at both sides of \llg{the} equality in~\eqref{eq:xinverse}, we obtain that $P(x)$ is palindromic.

\llg{Now we prove that $a_j$ in~\eqref{eq:px} and $b_j$ in~\eqref{eq:qx} must be either $0$ or $1$ by contradiction. Assuming otherwise, denote $l$ as the smallest integer such that $a_l \notin \{0,1\}$ or $b_l \notin \{0,1\}$. We first prove $l \leq \min \{k_1,k_2\}$. Without loss of generality, we assume $k_1 \leq k_2$. Assuming $l>k_1$, by the definition of $l$, we have $a_0,a_1\ldots,a_{k_1} \in \{0,1\}$, $b_0,b_1\ldots,b_{l-1} \in \{0,1\}$, and $b_l \notin \{0,1\}$. Comparing the coefficients of degree $l$ at both sides of equality in~\eqref{eq:ptimesq} yields
\begin{IEEEeqnarray}{rCl}
a_0b_l+a_1b_{l-1}+\cdots+a_{k_1}b_{l-k_1}=1.
\label{eq:llllg}
\end{IEEEeqnarray}
Note that $a_0=b_0=1$, then by~\eqref{eq:llllg} we have $b_l=1- (a_1b_{l-1}+\cdots+a_{k_1}b_{l-k_1})$. Recall that $a_1\ldots,a_{k_1} \in \{0,1\}$, $b_0,b_1\ldots,b_{l-1} \in \{0,1\}$, then $a_1b_{l-1}+\cdots+a_{k_1}b_{l-k_1}$ must be a nonnegative integer. Hence $b_l$ must also be an integer, and no larger than $1$. Since $b_l$ is nonnegative, $b_l$ must be either $0$ or $1$. This contradicts  the fact $b_l \notin \{0,1\}$. Hence $l \leq k_1$.

Based on this fact, we can rewrite~\eqref{eq:llllg} as
\begin{IEEEeqnarray}{rCl}
a_0b_l+a_1b_{l-1}+\cdots+a_lb_0=1.
\label{eq:llllg2}
\end{IEEEeqnarray}
Since $a_0=b_0=1$, ~\eqref{eq:llllg2} reduces to $a_l+b_l=1-(a_1b_{l-1}+\cdots+a_{l-1}b_{1})$. Recall that $a_0,a_1\ldots,a_{l-1} \in \{0,1\}$, $b_0,b_1\ldots,b_{l-1} \in \{0,1\}$, then $a_1b_{l-1}+\cdots+a_{l-1}b_{1}$ must be a nonnegative integer. Hence $a_l+b_l$ must also be an integer, and no larger than $1$. Since $a_l+b_l$ is nonnegative, $a_l+b_l$ must be either $0$ or $1$. By the definition of $l$, between the conditions $a_l \notin \{0,1\}$ and $b_l \notin \{0,1\}$, at least one of them must be true. In either case,  $a_l+b_l$ must be $1$, and $0<a_l<1$ and $0<b_l<1$. Hence $0<a_lb_l<1$.

We further compare the coefficients of degree $k_1$ at both sides of \llg{the} equality in~\eqref{eq:ptimesq}, and yield
\begin{IEEEeqnarray}{rCl}
a_{k_1}b_{0}+a_{k_1-1}b_{1}+\cdots+a_{k_l-l}b_{l}+\cdots+a_0b_{k_1}=1. \label{eq:lhscont}
\end{IEEEeqnarray}
By the palindrome of $P(x)$, we have $a_{k_1}=a_0=1$ and $a_l=a_{k_1-l}$. 
Then the LHS of~\eqref{eq:lhscont} satisfies
\begin{IEEEeqnarray}{rCl}
a_{k_1}b_{0}+a_{k_1-1}b_{1}+\cdots+a_{k_l-l}b_{l}+\cdots +a_0b_{k_1}&=&1+a_{k_1-1}b_{1}+\cdots+a_{l}b_{l}+\cdots +b_{k_1}>1. \nonumber \\
\end{IEEEeqnarray}
This contradicts the RHS of~\eqref{eq:lhscont}. Hence $a_l$ must be $0$ or $1$, and $b_l$ must also be $0$ or $1$. The proof is concluded.}
\end{IEEEproof}

\llg{Based on Proposition~\ref{prop:decompunifdisc}, we obtain the following corollary.}
\begin{corollary}
\label{cor:disunif}
\llg{Consider} three positive integers $k_1$, $k$, and $n$ with $(k_1+1)|(k+1)=n$. Let $U_{D_1}$ and $U_{D_2}$ be independent random variables whose support are the \llg{exact} $[k_1]$ and \llg{the} subset of $[k-k_1]$, respectively. Then the sum of $U_{D_1}$ and $U_{D_2}$, i.e., $U_D=U_{D_1}+U_{D_2}$, is uniformly distributed over the support $[k]$, if and only if $U_{D_1}$ is uniformly distributed over the support $[k_1]$, and $U_{D_2}$ is uniformly distributed over the support $\{j(k_1+1)|j \in [n-1]\}$.
\end{corollary}
\begin{IEEEproof}
By Proposition~\ref{prop:decompunifdisc} we only need to show that the support of \llg{$U_{D_2}$ is $\{j(k_1+1)|j \in [n-1]\}$}. Since $P_{U_D}(0)=P_{U_{D_1}}(0)P_{U_{D_2}}(0)$, Proposition~\ref{prop:decompunifdisc} further indicates $\forall l \in [k]$, there must exist a unique representation $l=l_1+l_2$, where $l_1 \in \mathsf{supp} U_{D_1}$ and $l_2 \in \mathsf{supp} U_{D_2}$, otherwise the uniformity of $U_D$ would be violated.

If there \llg{exist} two \llg{neighbor points} $r_1, r_2 \in \mathsf{supp} U_{D_2}$ satisfying $r_2-r_1<k_1+1$, then there are two representations $r_2=0+r_2=(r_2-r_1)+r_1$ where $r_2-r_1 \in [k_1]$, thus this violates the requirement of \llg{the} unique representation \llg{as mentioned above}. On the other hand, \llg{if there exist $r_1, r_2 \in \mathsf{supp} U_{D_2}$ satisfying $r_2-r_1 > k_1+1$, then for $r_1+k_1+1$ there is no representation from support of $U_{D_1}$ and $U_{D_2}$. Hence the distance of the neighbor points at the support of $U_{D_2}$ must be $k_1+1$.} Since $0 \in \mathsf{supp} U_{D_2}$ \llg{and $k-k_1 \in \mathsf{supp} U_{D_2}$}, the proof is concluded. 
\end{IEEEproof}

\begin{figure}[htbp]
\subfigure[a discrete distribution]{   
\begin{minipage}{12cm}
\centering    
\includegraphics[scale=0.7]{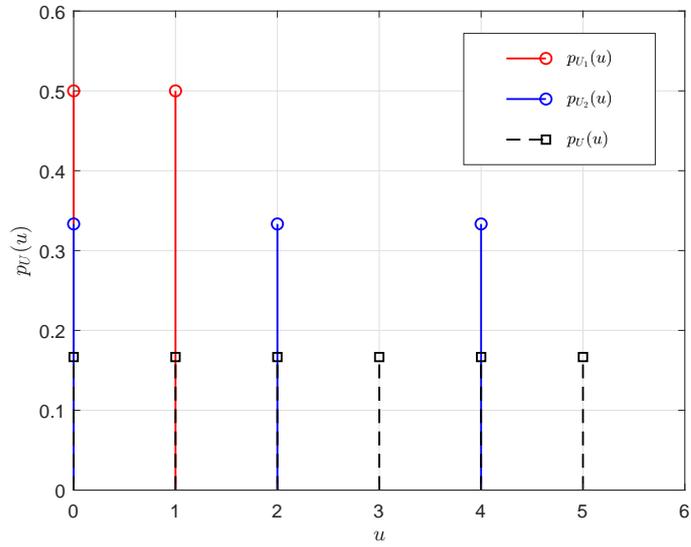}  
\end{minipage}
}
\subfigure[a truncated geometric distribution]{ 
\begin{minipage}{12cm}
\centering    
\includegraphics[scale=0.7]{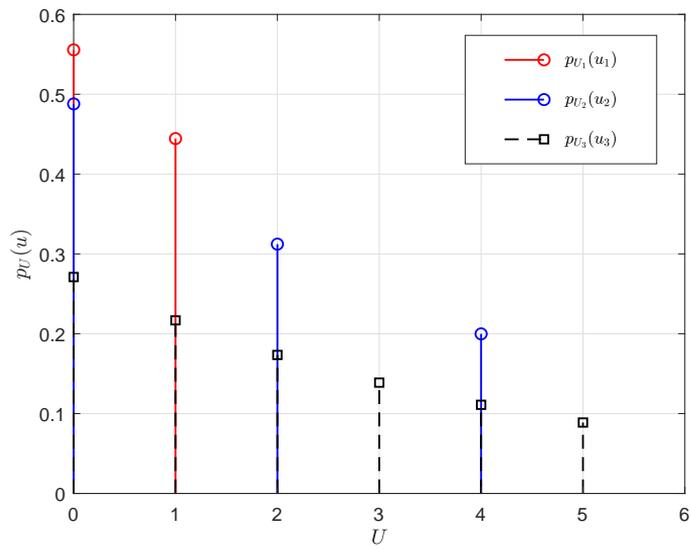}
\end{minipage}
}
\caption{Decomposition of a discrete uniform and a truncated geometric distribution with $k_1=1$ and $k_2=4$.}
\label{fig:02}
\end{figure}

\subsubsection{Decomposition of a geometric distributed random variable}
The following proposition presents the result on the decomposition of a geometric distributed random variable.
\begin{proposition}
\label{discrgeometdistr}
Given $\lambda,~\lambda_1 \in (0,1)$ with $\lambda<\lambda_1$\llg{, consider} a random variable $U_D$ following a geometric distribution with parameter $\lambda$, i.e.,
\begin{IEEEeqnarray}{rCl}
p_{U_D}(k)=\lambda(1-\lambda)^k, \quad \forall k \in \mathbb{N}.
\end{IEEEeqnarray} 
$U$ can be decomposed as a sum of two independent random variables, i.e., 
\begin{IEEEeqnarray}{c}
U_D = U_{D_1}+U_{D_2},
\end{IEEEeqnarray}
where $ U_{D_1}$ follows a geometric distribution with parameter $\lambda_1$, and $U_{D_2}$ follows the following distribution:
\begin{IEEEeqnarray}{c}
 p_{U_{D_2}}(k)=\frac{\lambda}{\lambda_1} \frac{1-\lambda_1}{1-\lambda}  \delta (k)+ \frac{\lambda_1-\lambda}{\lambda_1  (1-\lambda)} \lambda(1-\lambda)^k, \quad \forall k \in \mathbb{N}.
\end{IEEEeqnarray} 
\end{proposition}
\begin{IEEEproof}
The c.f. of $U_{D_1}$ is given by
\begin{IEEEeqnarray}{c}
\phi_{U_{D_1}}(t) = \frac{\lambda_1}{1-(1-\lambda_1)e^{it}}.
\end{IEEEeqnarray}
Similarly, for $ U_{D_2}$ we have 
\begin{IEEEeqnarray}{rCl}
\phi_{U_{D_2}}(t) &=& \frac{\lambda}{\lambda_1} \frac{1-\lambda_1}{1-\lambda}+ \frac{\lambda_1-\lambda}{\lambda_1  (1-\lambda)} \frac{\lambda}{1-(1-\lambda)e^{it}} \nonumber \\
&=& \frac{\lambda}{\lambda_1} \frac{1-(1-\lambda_1)e^{it}}{1-(1-\lambda)e^{it}}.
\end{IEEEeqnarray}
The proof is concluded by the fact
\begin{IEEEeqnarray}{rCl}
\phi_{U_{D_1}}(t)  \phi_{U_{D_2}}(t) &=& \frac{\lambda_1}{1-(1-\lambda_1)e^{it}}  \frac{\lambda}{\lambda_1} \frac{1-(1-\lambda_1)e^{it}}{1-(1-\lambda)e^{it}}  \\ 
 &=& \frac{\lambda}{1-(1-\lambda)e^{it}}\\
 &=&\phi_{U_{D}}(t) .
\end{IEEEeqnarray}
\end{IEEEproof}

\subsubsection{Decomposition of a truncated geometric distributed random variable}
Following the same spirit as in Proposition~\ref{prop:decompunifdisc}, the following proposition \llg{provides} a necessary condition on the distributions of the decomposed independent random variables from a truncated geometric distribution
\begin{proposition}
\label{prop:decompdistrunexporv}
\llg{Consider} three positive integers $k_1$,~$k_2$, and $k$ with $k=k_1+k_2$, and a real number $\lambda \in (0,1)$. Let $U_{D_1}$ and $U_{D_2}$ be independent random variables whose support are the subsets of $[k_1]$ and $[k_2]$, respectively. If the sum of $U_{D_1}$ and $U_{D_2}$, i.e., $U_D=U_{D_1}+U_{D_2}$, follows a truncated geometric distribution with parameter $\lambda$ over support $[k]$, then $\forall j \in \mathsf{supp} U_{D_n}$,  $p_{U_{D_n}}(j) = p_{U_{D_n}}(0) (1-\lambda)^{j} $,$\,\,n=1,2$.
\end{proposition}
\begin{IEEEproof}
Define two polynomials $P(x)$ and $Q(x)$ as in~\eqref{eq:px} and~\eqref{eq:qx}. Then we have 
\begin{IEEEeqnarray}{rCl}
P(x)  Q(x)=\sum_{j \in [k]}((1-\lambda)x)^j. \label{eq:ptimesq22}
\end{IEEEeqnarray}
Now we rewrite $P(x)$ and $Q(x)$ as 
\begin{IEEEeqnarray}{rCl}
P(x)=\sum_{j \in [k_1]}\frac{a_j}{(1-\lambda)^j}((1-\lambda)x)^j, 
\end{IEEEeqnarray}
and 
\begin{IEEEeqnarray}{rCl}
Q(x)=\sum_{j \in [k_2]}\frac{b_j}{(1-\lambda)^j}((1-\lambda)x)^j. 
\end{IEEEeqnarray}
Following the similar arguments as in the proof of~\llg{Proposition~\ref{prop:decompunifdisc}}, we can show that $\frac{a_j}{(1-\lambda)^j}$'s (and $\frac{b_j}{(1-\lambda)^j}$'s) are either $0$ or $1$. The proof is concluded.
\end{IEEEproof}
Based on Proposition~\ref{prop:decompdistrunexporv}, we present the following corollary that is useful in our paper. The proof follows the same arguments as in the proof of Corollary~\ref{cor:disunif}. \llg{We omit it here}.
\begin{corollary}
\label{cor:distrunexp}
\llg{Consider} three positive integers $k_1$, $k$, and $n$ with $(k_1+1)|(k+1)=n$, and a real number $\lambda \in (0,1)$. Let $U_{D_1}$ and $U_{D_2}$ be independent random variables whose support are the \llg{exact} $[k_1]$ and the subset of $[k-k_1]$, respectively. Then the sum of $U_{D_1}$ and $U_{D_2}$, i.e., $U_D=U_{D_1}+U_{D_2}$, follows a truncated geometric distribution with parameter $\lambda$ over the support $[k]$, if and only if $U_{D_1}$ follows a truncated geometric distribution with parameter $\lambda$ over the support $[k_1]$, and $U_{D_2}$ also follows a truncated geometric distributed with parameter $1-(1-\lambda)^{k_1+1}$ over the support $\{j(k_1+1)|j \in [n-1]\}$, i.e.,
\begin{IEEEeqnarray}{rCl}
p_{U_{D_1}} (j)= \frac{\lambda}{1-(1-\lambda)^{k_1+1}}  (1-\lambda)^{j}, \quad j \in [k_1],
\end{IEEEeqnarray}
and
\begin{IEEEeqnarray}{rCl}
p_{U_{D_2}} (j(k_1+1))= \frac{1-(1-\lambda)^{k_1+1}}{1-(1-\lambda)^{k+1}}  (1-\lambda)^{j(k_1+1)}, \quad j \in [n-1].
\end{IEEEeqnarray}
\end{corollary}
 Figure~\ref{fig:02} shows an example \llg{of} the \llg{decomposition} of a uniform distribution in Corollary~\ref{cor:disunif} and a truncated geometric distribution with parameter $\lambda=0.2$ in Corollary~\ref{cor:distrunexp}. These two distributions hold similar characteristics as in previous mentioned continuous uniform and truncated distributions, \llg{both of which} can be decomposed as a sum of a `contracted' version of themselves and an equi-spaced discrete distribution.

\section{Capacity Results}
\label{sec:caparesults}
In this section, we \llg{present} bounds \llg{on the sum-capacity} of the peak- or/and average-intensity constrained MAC based on the \llg{decomposition} results in Sec.~\ref{sec:decomprv}. Before presenting the bounds, we first give an auxiliary result on \llg{the} capacity of a single-user optical intensity channel \llg{derived by} discrete equi-spaced \llg{inputs}.
\subsection{An Auxiliary Capacity Lower Bound of a Single-User \llg{ Optical Intensity} Channel}
\llg{Exact characterization of the capacity of the optical intensity channel with a peak-limited input is still an open problem, but} it is \llg{known} that the capacity-achieving input distribution is discrete with a finite support~\cite{smith71_1,chanhranilovickschischang05_1,sharma2010transition,dytso2019capacity}. At moderate and low SNR \llg{regimes}, existing results show that the achievable rates by \llg{certain discrete input distributions} are close to the channel capacity~\cite{chaaban2020capacity,chaaban2017capacity}. Based on this observation, we present {several} closed-form lower \llg{bounds} on single-user channel capacity by applying a recent result in~\cite{melbourne2022differential} on the entropy of Gaussian-mixture \llg{distribution}.

For a single-user \llg{optical} intensity channel in~\eqref{eq:singlechannelmodel}, consider a discrete random variable $X_D$. The support of $X_D$ is an equi-spaced discrete set $\Pi=\{ 2lj|j\in [K-1]\}$, where \llg{the} integer $K$ denotes its cardinality with $K \geq 2$, and \llg{the} points are spaced $2l$ apart with $l>0$. Assume $X_D$ is an admissible input, then the achievable rate $\mathsf{I}(X_D;Y)$ can serve as a capacity lower bound, which can be expanded as 
\begin{IEEEeqnarray}{rCl}
\label{eq:iudy}
\mathsf{I}(X_D;Y) = \mathsf{H}(X_D)-\mathsf{H}(X_D|Y).
\end{IEEEeqnarray}
Now we lower bound the conditional entropy term $\mathsf{H}(X_D|Y)$ in~\eqref{eq:iudy} by the following lemma.
\begin{lemma}[{\cite[Cor.~$9$]{melbourne2022differential}}]
\label{lemma15}
Consider a discrete random variable $U_D$ where the space between points in the support is no less than $2l$ with $l>0$, and an independent Gaussian random variable $Z\sim \mathcal{N}(0,\sigma^2)$, then
\begin{IEEEeqnarray}{rCl}
\mathsf{H}(X_D|X_D+Z) \leq R\left(\frac{\sigma}{l}\right), 
\end{IEEEeqnarray}
where \llg{the} function 
\begin{IEEEeqnarray}{rCl}
R\left(u\right) =2\mathcal{Q}\left(\frac{1}{u}\right)\left(2+\frac{1}{2u^2}+\log (5+2u^2)+\log \left(1+\sqrt{\frac{9\pi}{2}}u\right)\right),\,\, u>0.
\label{eq:ru}
\end{IEEEeqnarray}
\end{lemma}
Substituting~\eqref{eq:ru} into~\eqref{eq:iudy}, we have 
\begin{IEEEeqnarray}{rCl}
\label{eq:lowbndud}
\mathsf{I}(X_D;Y) \geq \mathsf{H}(X_D)-R\left(\frac{\sigma}{l}\right).
\end{IEEEeqnarray}
When $l$ and $K$ are fixed, to get a good lower bound on $\mathsf{I}(X_D;Y)$ we only need to maximize $\mathsf{H}(X_D)$ over all admissible \llg{inputs} in \llg{the} discrete set $\Pi$. In the following, we give \llg{the} lower capacity bounds when the input is subject to \llg{the} peak- or/and average-\llg{intensity constraints}.
\begin{proposition}
\label{proplowerbndsingle}
When $K$ is finite, \llg{in the peak-intensity} constrained channel, 
\begin{IEEEeqnarray}{rCl}
\mathsf{I}(X_D;Y) \geq \log (K+1) - R(2K \sigma); \label{eq:peaklbnd}
\end{IEEEeqnarray}
\llg{in the average-intensity} constrained channel,
\begin{IEEEeqnarray}{rCl}
\mathsf{I}(X_D;Y) &\geq& \sup_{t \in (0,1)} \bigg\{ \log \frac{1-{(1-t)}^{K+1}}{t} - \left(K+\frac{1}{t}-\frac{K+1}{1-(1-t)^{K+1}}\right) \log (1-t) \nonumber \\
&&\hspace{3.2cm}- R\left(2 \left(K+\frac{1}{t}-\frac{K+1}{1-(1-t)^{K+1}}\right)  \frac{\sigma}{\alpha}\right)\bigg\};
\label{eq:averagelowbnd}
\end{IEEEeqnarray}
in the peak- and average-intensity constrained  channel,
\begin{IEEEeqnarray}{rCl}
\mathsf{I}(X_D;Y) &\geq& \sup_{t \in (0,\eta)} \bigg\{ \log \frac{1-{(1-t)}^{K+1}}{t} - \left(K+\frac{1}{t}-\frac{K+1}{1-(1-t)^{K+1}}\right) \log (1-t) \nonumber \\
&&\hspace{3.2cm}- R\left(2 \left(K+\frac{1}{t}-\frac{K+1}{1-(1-t)^{K+1}}\right)  \frac{\sigma}{\alpha}\right) \bigg\},
\label{eq:peakavermmm}
\end{IEEEeqnarray}
where $\eta$ is the unique solution to 
\begin{IEEEeqnarray}{rCl}
\frac{1}{\eta}-\frac{K+1}{1-(1-\eta)^{K+1}}+(1-\alpha) K=0.
\end{IEEEeqnarray}
Moreover, when $K=\infty$, \llg{in the average-intensity} constrained channel, 
\begin{IEEEeqnarray}{rCl}
\mathsf{I}(X_D;Y) \geq \sup_{l\in(0,\infty)}  \left\{\log\left(1+\frac{\alpha}{2l}\right)+\frac{\alpha}{2l}\log\left(1+\frac{2l}{\alpha}\right) - R\left(\frac{\sigma}{l}\right) \right\}  . \label{eq:maxgeomtric}
\end{IEEEeqnarray}
\end{proposition}
\llg{
\begin{IEEEproof}
See Appendix~\ref{appproplowerbndsingle}.
\end{IEEEproof}
}
\subsection{Peak-Intensity Constrained MAC }
\label{sec:peak}
In the peak-intensity constrained MAC, without loss of generality, we assume $\mathsf{A}_1 \leq \mathsf{A}_2$ and $\mathsf{A}_1 + \mathsf{A}_2 =1$. The results in this section can be directly \llg{applied} to the peak-power limited Gaussian MACs~\cite{mamandipoor2014capacity}. We first present \llg{an} upper bound on the sum-capacity based on the capacity result of \llg{the} single-user channel with peak-intensity constraint in~\cite{mckellips04_1,zhou19_1}.

\begin{proposition}[{\cite[eq.~$(1)$]{mckellips04_1},~\cite[Prop.~$5$]{zhou19_1}}]
The sum-capacity of \llg{the} peak-intensity constrained MAC is upper bounded by
\begin{IEEEeqnarray}{c}
\label{upperbnd}
\mathsf{C}_{\textnormal{p-sum}} \leq \min \left\{\frac{1}{2}\log \left(1+\frac{1}{4\sigma^2} \right),\,\log \left(1+ \frac{1}{\sqrt{2\pi e}\sigma}\right) \right\}.
\end{IEEEeqnarray}
\end{proposition}
Now we present \llg{a} lower bound by applying the \llg{result} in Lemma~\ref{lem1}. 
\begin{theorem}
\label{prop:lowbnd1}
The sum-capacity of \llg{the} peak-intensity constrained MAC is lower bounded by
\begin{IEEEeqnarray}{c}
\label{lowerbnd}
\mathsf{C}_{\textnormal{p-sum}} \geq  \frac{1}{2}\log \left( 1+ \frac{1}{2\pi e \sigma^2} \right).
\end{IEEEeqnarray}
\end{theorem}
\begin{IEEEproof}
Let $a=\mathsf{A}_1$ in Lemma~\ref{lem1}. By the assumption $\mathsf{A}_1+\mathsf{A}_2=1$, it is direct to see the support of $U_1$ and $U_2$ are \llg{the} subsets of $[0,\mathsf{A_1}]$ and $[0,\mathsf{A_2}]$, respectively. Also, by Lemma~\ref{lem1},  $U_1$ and $U_2$ are independent. Let $X_1 = U_1 $, and $X_2 =U_2 $, then $(X_1,X_2)$ \llg{forms} an admissible input pair.
By the EPI~\cite{coverthomas06_1} \llg{and the fact} that $X_1+X_2$ is uniformly distributed \llg{over} $[0,1]$, we have  
\begin{IEEEeqnarray}{rCl}
I(X_1+X_2;Y) &\geq&  \frac{1}{2}\log \left( 1+ e^{2h(X_1+X_2)-2h(Z)} \right) \\
             &=& \frac{1}{2}\log \left( 1+ \frac{1}{2\pi e \sigma^2} \right).
\end{IEEEeqnarray}
\end{IEEEproof}
Combined with~\eqref{upperbnd} and~\eqref{lowerbnd}, and let $\sigma \rightarrow 0$, we characterize the high-SNR asymptotic capacity. 
\begin{corollary}
The high-SNR asymptotic capacity of the peak-intensity constrained MAC is given by

\begin{IEEEeqnarray}{c}
\lim_{\sigma \rightarrow 0}\{\mathsf{C}_{\textnormal{p-sum}}+\log{\sigma}\} =  -\frac{1}{2}\log ({2\pi e}).
\end{IEEEeqnarray}
\end{corollary}

\llg{For some values of $\mathsf{A}_1$ (or $\mathsf{A}_2$), the distribution of $X_1$ (or $X_2$) in the proof of Theorem~\ref{prop:lowbnd1} may be singular.} Proposition~\ref{prop1} shows that the uniform distribution over $[0,1]$ can be explicitly decomposed as a sum of a `contracted' uniform distribution over $\left[0,\frac{1}{k}\right]$ with \llg{the} integer $k \geq 2$,  and a $k$-point equi-spaced discrete uniform distribution. Based on this result, the following proposition presents another simple lower bound.
\begin{proposition}
The sum-capacity of \llg{the} peak-intensity constrained MAC is lower bounded by
\begin{IEEEeqnarray}{c}
\label{lowerbndbnd}
\mathsf{C}_{\textnormal{p-sum}} \geq  \frac{1}{2}\log \left( 1+ \frac{\left( \left\lfloor\frac{1}{\mathsf{A}_1}\right\rfloor\frac{1}{\left\lceil \frac{1}{\mathsf{A}_1} \right\rceil}\right)^2 }{2\pi e \sigma^2} \right).
\end{IEEEeqnarray}
\end{proposition}
\begin{IEEEproof}
 Since $\mathsf{A}_1 \in \left(0,\frac{1}{2}\right]$, \llg{we have $\mathsf{A}_1 \in \left[\frac{1}{k},\frac{1}{k-1}\right)$, where \llg{the} integer $k= \left\lceil\frac{1}{\mathsf{A}_1}\right\rceil$}. Let $X_1$ \llg{follow} \llg{a} uniform distribution over the support $\left[0,\frac{1}{k}\right]$, and it is direct to see $X_1$ satisfies the peak-intensity constraint.

Let $X_2$ follow \llg{a} $\left\lfloor \frac{1}{\mathsf{A}_1} \right\rfloor$-point uniform distribution over the support $\left\{\frac{j}{k}|j\in \left[\left\lfloor \frac{1}{\mathsf{A}_1} \right\rfloor-1\right]\right\}$. We only need to \llg{verify} $X_2$ satisfies the peak-intensity constraint. \llg{We consider two different cases}. When $\mathsf{A}_1 = \frac{1}{k}$, then $ \left\lfloor \frac{1}{\mathsf{A}_1} \right\rfloor = \left\lceil \frac{1}{\mathsf{A}_1} \right\rceil =k $, and the \llg{largest value of} mass point in the support of $X_2$ is $\frac{k-1}{k}=1-\mathsf{A}_1=\mathsf{A}_2$. When $\mathsf{A}_1 \in (\frac{1}{k},\frac{1}{k-1})$, then $ \left\lfloor \frac{1}{\mathsf{A}_1} \right\rfloor = \left\lceil \frac{1}{\mathsf{A}_1} \right\rceil-1 =k-1 $, and \llg{the largest value of} mass point in the support of $X_2$ is $\frac{k-2}{k}=1-\frac{2}{k}<1-\mathsf{A}_1=\mathsf{A}_2$.  Hence $X_2$ satisfies the peak-intensity constraint in both cases, and $(X_1,X_2)$ forms an admissible input pair. By Proposition~\ref{prop1}, $X_1+X_2$ follows a uniform distribution over $\left[0,\left\lfloor\frac{1}{\mathsf{A}_1}\right\rfloor\frac{1}{\left\lceil \frac{1}{\mathsf{A}_1} \right\rceil}\right]$. The proof is concluded by applying the EPI.
\end{IEEEproof}

We further present a lower bound based on an equi-spaced discrete input and by applying~\eqref{eq:peaklbnd} in Proposition~\ref{proplowerbndsingle}.
\begin{proposition}
The sum-capacity of \llg{the} peak-intensity constrained MAC is lower bounded by
\begin{IEEEeqnarray}{rCl}
\mathsf{C}_{\textnormal{p-sum}} \geq  \log \left(n   \left\lfloor\frac{1}{\mathsf{A}_1}\right\rfloor\right) - R\left(\frac{2n  \sigma}{\mathsf{A}_1}\right),
\label{eq:llowbndw}
\end{IEEEeqnarray}
where \llg{the} integer $n \geq 2$, and is a free parameter. 
\end{proposition}
\begin{IEEEproof}
Let $X_1$ be uniformly distributed over the support $\left\{\frac{j \mathsf{A}_1}{n}\big|j \in [n-1]\right\}$, and $X_2$ \llg{be} uniformly distributed over the support $\left\{j \mathsf{A}_1\big|j\in \left[\left\lfloor \frac{1}{\mathsf{A}_1} \right\rfloor-1 \right]\right\}$. It is direct to verify $(X_1,X_2)$ forms an admissible input pair. Then by Proposition~\ref{prop:decompunifdisc}, we have $X_1+X_2$ follows an $\left(n    \left\lfloor\frac{1}{\mathsf{A}_1}\right\rfloor   \right)$-point uniform distribution over the support $\left\{\frac{j \mathsf{A}_1}{n}\big| j \in \left[ n \left\lfloor \frac{1}{\mathsf{A}_1} \right\rfloor-1 \right]  \right\}$. The proof is concluded by applying~\eqref{eq:peaklbnd} in Proposition~\ref{proplowerbndsingle}.
\end{IEEEproof}

\begin{figure}[h]
	\centering
	\resizebox{12cm}{!}{\includegraphics{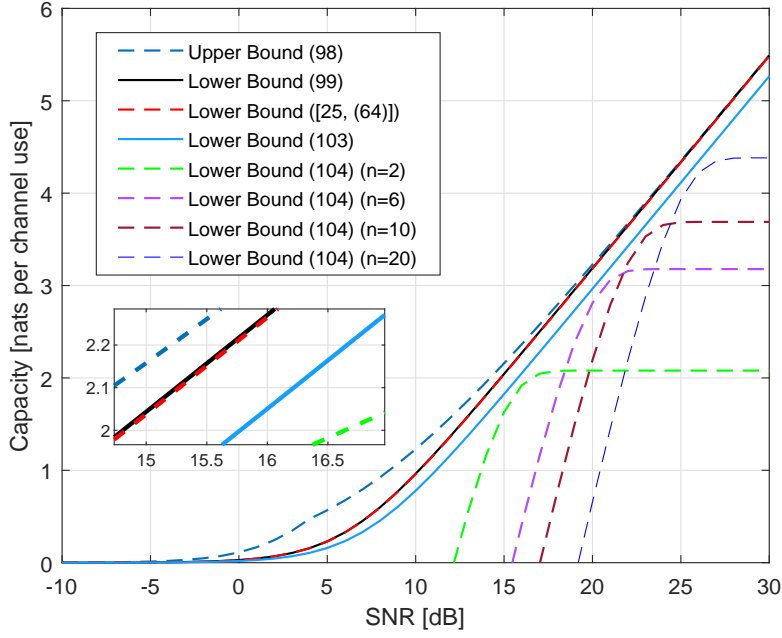}}
	\centering \caption{A two-user peak-intensity constrained MAC channel ($\mathsf{A}_1 = 0.3$).}
  \label{fig:1}
\end{figure}

\llg{Figure}~\ref{fig:1} depicts \llg{the} derived sum-capacity bounds by one example \llg{with} $\mathsf{A}_1=0.3$ and $\mathsf{A}_2=0.7$. At moderate and high SNRs, the lower bound in~\eqref{lowerbnd} gives the best approximation \llg{to the sum-capacity}. We can also directly see that the lower bounds in~\eqref{lowerbndbnd} and~\eqref{eq:llowbndw} are also fairly close to the capacity at moderate and high SNR. 

\begin{figure}[h]
	\centering
	\resizebox{12cm}{!}{\includegraphics{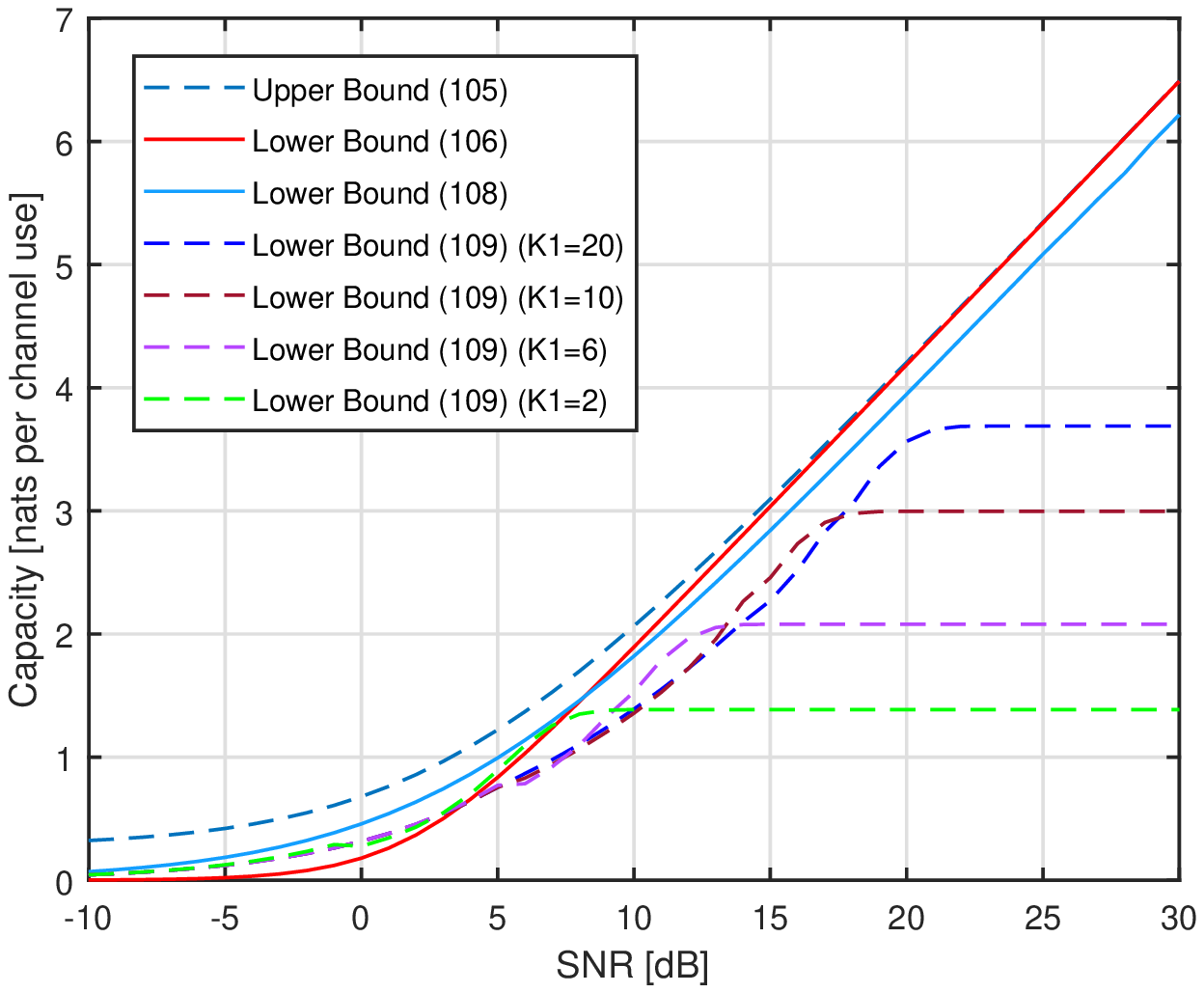}}
	\centering \caption{A two-user average-intensity constrained MAC channel
    ($\mathsf{E}_1 = 0.44$).}
  \label{fig:2}
\end{figure}

\subsection{Average-Intensity Constrained MAC }
Similar \llg{to} Section~\ref{sec:peak}, the following upper bound \llg{on} sum-capacity is based on the result of single-user channel with \llg{the} average-intensity constraint in~\cite{hranilovickschischang04_1,zhou19_1}.

\begin{proposition}[{\cite[eq.~($20$)]{hranilovickschischang04_1},~\cite[Prop.~$1$]{zhou19_1}}]
The sum-capacity of \llg{the} average-intensity constrained MAC is upper bounded by\footnote{This upper bound is implicitly given in~\cite[eq. (20)]{hranilovickschischang04_1} by letting  $V(\Upsilon_1)=1$, $N=1$, and $\sqrt{T}P =1$. See footnote~$2$ in~\cite{zhou19_1} for more details.}
\begin{IEEEeqnarray}{c}
\label{upperbnda}
\mathsf{C}_{\textnormal{a-sum}} \leq \frac{1}{2}\log\left(\frac{e}{2\pi}\left(\frac{1}{\sigma}+2\right)^2 \right).
\end{IEEEeqnarray}
\end{proposition}
In the following, we recover a lower bound in~\cite{zhou19_1} by applying the results in Proposition~\ref{lem2}. 
\begin{theorem}[{\cite[eq.~($26$)]{lapidothmoserwigger09_7},~\cite[Prop.~$2$]{zhou19_1}}]
The sum-capacity of \llg{the} average-intensity constrained MAC is lower bounded by
\begin{IEEEeqnarray}{c}
\label{lowerbndb}
\mathsf{C}_{\textnormal{a-sum}} \geq  \frac{1}{2}\log \left( 1+ \frac{e}{2\pi \sigma^2} \right).
\end{IEEEeqnarray}
\end{theorem}

\begin{IEEEproof}
Let $a=\mathsf{A}_1$ in Proposition~\ref{lem2}, and then we set $X_1 = U_1 $, and $X_2 =U_2 $. It is direct to verify $(X_1,X_2)$ forms an admissible input pair. The proof can be completed by following the same \llg{arguments} as in the proof of Proposition~\ref{prop:lowbnd1}.
\end{IEEEproof}

Combined with~\eqref{upperbnda} and~\eqref{lowerbndb}, and let $\sigma \rightarrow 0$, we characterize the high-SNR asymptotic \llg{sum-capacity}.
\begin{corollary}
The high-SNR asymptotic \llg{sum-capacity} of \llg{the} average-intensity constrained MAC is given by

\begin{IEEEeqnarray}{c}
\lim_{\sigma \rightarrow 0}\{\mathsf{C}_{\textnormal{a-sum}}+\log{\sigma}\} =  \frac{1}{2}\log \frac{e}{2\pi}.
\end{IEEEeqnarray}
\end{corollary}

In the following, we present a lower bound based on an equi-spaced discrete input and by applying~\eqref{eq:maxgeomtric} and~\eqref{eq:averagelowbnd} in Proposition~\ref{proplowerbndsingle}.
\begin{proposition}
The sum-capacity of average-intensity constrained MAC is lower bounded by
\begin{IEEEeqnarray}{c}
\mathsf{C}_{\textnormal{a-sum}} \geq  \sup_{l\in(0,\infty)} \left\{ \log\left(1+\frac{1}{2l}\right)+\frac{1}{2l}\log\left(1+{2l}\right) - R\left(\frac{\sigma}{l}\right)\right\},
\label{eq:geometric}
\end{IEEEeqnarray}
and
\begin{IEEEeqnarray}{rCl}
\mathsf{C}_{\textnormal{a-sum}}  &\geq& \sup_{t \in (0,1)} \bigg\{ \log \frac{1-{(1-t)}^{K}}{t} - \left(K-1+\frac{1}{t}-\frac{K}{1-(1-t)^{K}}\right) \log (1-t) \nonumber \\
&&\hspace{3.2cm}- R\left(2 \left(K_1-1+\frac{1}{t}-\frac{K_1}{1-(1-t)^{K_1}}\right)  \frac{\sigma}{\mathsf{E}_1}\right)\bigg\},
\label{eq:truncageometric}
\end{IEEEeqnarray}
where $K= K_1  n$ with \llg{the} integer $K_1 \geq 2$ and $n$ satisfying
\begin{IEEEeqnarray}{rCl}
n=\argmax_{m \in \mathbb{N}}\left\{ \frac{1}{t}-\frac{m K_1}{1-(1-t)^{m  K_1}}+m  K_1-1\leq \frac{1}{2l} \right\},
\label{eq:etaeq2}
\end{IEEEeqnarray}
with $l$ being the unique solution to 
\begin{IEEEeqnarray}{rCl}
\frac{1}{t}-\frac{K_1}{1-(1-t)^{K_1}}+K_1-1=\frac{\mathsf{E}_1}{2l}.
\label{eq:etaeq}
\end{IEEEeqnarray}

\end{proposition}
\begin{IEEEproof}
\llg{We first prove~\eqref{eq:geometric}.} Given an $l \in (0,\infty)$, consider two random \llg{variables} $U_{D_1}$ and $U_{D}$ following \llg{geometric distributions with parameters $\frac{2l}{2l+\mathsf{E}_1}$ and $\frac{2l}{2l+1}$, respectively, i.e.,}
\begin{IEEEeqnarray}{rCl}
p_{U_{D_1}}(j) = \frac{2l}{(2l+\mathsf{E}_1)^{j+1}}  {\mathsf{E}_1}^{j}, \quad \forall j \in \mathbb{N}.
\end{IEEEeqnarray}
and
\begin{IEEEeqnarray}{rCl}
p_{U_{D}}(j) = \frac{2l}{(2l+1)^{j+1}}, \quad \forall j \in \mathbb{N}.
\end{IEEEeqnarray}

It is direct to verify $\mathbb{E}[U_{D_1}] = \mathsf{E}_1$, \llg{and $\mathbb{E}[U_{D}] = 1$}. Then by Proposition~\ref{discrgeometdistr}, there exists a random \llg{variable} $U_{D_2}$, independent of $U_{D_1}$, such that $U_{D_2}=U_D-U_{D_1}$, and satisfies $\mathbb{E}[U_{D_2}]=\mathbb{E}[U_{D}]-\mathbb{E}[U_{D_1}]=1-\mathsf{E}_1=\mathsf{E}_2$. Let $X_1=U_{D_1}$ and $X_2=U_{D_2}$, then $(X_1,X_2)$ forms an admissible input \llg{pair}, and the proof is concluded by applying~\eqref{eq:maxgeomtric} in Proposition~\ref{proplowerbndsingle}.

To prove~\eqref{eq:truncageometric}, let $X_1$ \llg{follow} a $K_1$-point truncated geometric distribution: 
\begin{IEEEeqnarray}{rCl}
p_{X_1}(j) = \frac{\eta  (1-\eta)^{j} }{1-(1-\eta)^{K_1}}, \quad \forall j \in [K_1-1],
\end{IEEEeqnarray}
where $\eta$ is the unique solution to
\begin{IEEEeqnarray}{rCl}
\frac{1}{\eta}-\frac{K_1}{1-(1-\eta)^{K_1}}+K_1-1=\frac{\mathsf{E}_1}{2l}.
\end{IEEEeqnarray}
Then we have $\mathbb{E}[X_1]=\mathsf{E}_1$.
Further let $X_2$ follow \llg{an} $n$-point truncated geometric distribution:
\begin{IEEEeqnarray}{rCl}
p_{X_2} (j)= \frac{1-(1-\eta)^{K_1}}{1-(1-\eta)^{K}}  (1-\eta)^{j K_1}, \quad \forall j \in [n-1].
\end{IEEEeqnarray}
By Corollary~\ref{cor:distrunexp}, $ X_1+X_2$ follows an $(n  K_1)$-point truncated geometric distribution over \llg{the} support $\{ 2lj|j\in [n K_1-1]\}$\llg{, and
\begin{IEEEeqnarray}{rCl}
\mathbb{E}[X_1+X_2]=2l  \left(\frac{1}{t}-\frac{n K_1}{1-(1-t)^{n  K_1}}+n K_1-1\right).
\end{IEEEeqnarray}}
By~\eqref{eq:etaeq2} and~\eqref{eq:etaeq}, it is direct to verify $\mathbb{E}[X_2]=\mathbb{E}[X_1+X_2]-\mathbb{E}[X_1] \leq 1-\mathsf{E}_1=\mathsf{E}_2$. Hence $(X_1,X_2)$ forms an admissible input pair. The proof is concluded by applying~\eqref{eq:averagelowbnd} to $X_1+X_2$.
\end{IEEEproof}

\llg{Figure}~\ref{fig:2} plots \llg{the derived} capacity bounds by one example with $\mathsf{E}_1=0.44$ and $\mathsf{E}_2=0.56$. At high SNR, the lower bound in~\eqref{lowerbndb} asymptotically matches the upper bound in~\eqref{upperbnda}. Lower bounds in~\eqref{eq:geometric} and~\eqref{eq:truncageometric} give better approximations to \llg{the sum-capacity} at moderate and low SNR.

\subsection{Peak- and Average-Intensity Constrained MAC}

For the convenience of \llg{notations}, \llg{in the following, we define three functions:} 
\begin{IEEEeqnarray}{rCl}
f(c,v)= \frac{1}{v}- \frac{ce^{-cv}}{1-e^{-cv}},\quad c \in \left(\frac{1}{2},1\right],\; v \in (0,\infty);
\end{IEEEeqnarray}

\begin{IEEEeqnarray}{rCl}
g(c,v,m)= f(c,v)- f\left(\frac{c}{m},v\right),\quad c \in \left(\frac{1}{2},1\right],\; v \in (0,\infty),\;  m \in \mathbb{N}^+; \label{eq:gx}
\end{IEEEeqnarray}
and 
\begin{IEEEeqnarray}{rCl}
h(c,v)= 1- \frac{cve^{-cv}}{1-e^{-cv}}-\log \frac{v}{1-e^{-cv}}, \quad c \in \left(\frac{1}{2},1\right],\; v \in (0,\infty).
\end{IEEEeqnarray}
By basic calculus arguments, we can \llg{prove that} $f(c,v)$, $g(c,v,m)$, and $h(c,v)$ are \llg{all} monotonically increasing over $c$ and decreasing over $v$, and $g(c,v,m)$ monotonically increasing over $m$. The range of $f(c,v)$, $g(c,v,m)$, and $h(c,v)$ are $\left(0,\frac{c}{2}\right)$,~$\left(0,\frac{c}{2}\left(1-\frac{1}{m}\right)\right)$,~and $(-\infty,\log c)$, respectively.
\begin{remark}
Consider a truncated exponential random variable $U$ over $[0,c]$ with parameter $v$. Then we can directly get $\mathbb{E}[U]=f(c,v)$, and $\mathsf{h}(U)=h(c,v)$. By Proposition~\ref{prop9}, $U$ can be decomposed as a sum of two independent random variables $U_1$ and $U_2$, where $U_1$ is a truncated exponential random variable over $\left[0,\frac{c}{m}\right]$ with parameter $v$, and $U_2$ is a truncated geometrically distributed random variable over the support $\left\{\frac{j c}{m}|j \in [m-1] \right\}$ with parameter $1-e^{-\frac{vc}{m}}$. Then we have $\mathbb{E}[U_1]=f\left(\frac{c}{m},v\right)$, and $\mathbb{E}[U_2]=g(c,v,m)$.
\end{remark}

\llg{In the following, we present an upper bound on the sum-capacity which is based on the capacity result of the single-user channel with \llg{the} peak- and average-intensity constraints in~\cite{li2018miso}.}
\begin{proposition}[{\cite[Prop.~$6$]{li2018miso}}]
The sum-capacity of \llg{the} peak- and average-intensity constrained MAC is upper bounded by
\begin{IEEEeqnarray}{c}
\mathsf{C}_{\textnormal{ap-sum}} \leq \log \left(1+\frac{1}{\sqrt{2\pi e}\sigma}\frac{1-e^{-\eta}}{\eta}\right)+ \frac{\eta \sigma}{\sqrt{2\pi}}\left(1-e^{-\frac{1}{2\sigma^2}}\right)+\eta \alpha_{\textnormal{w}},
\label{eq:sum-upp}
\end{IEEEeqnarray}
where \llg{the free parameter $\eta>0$}, \llg{and $\alpha_{\textnormal{w}}$ is defined in~\eqref{eq:alphasum}}.
\end{proposition}

It is shown in~\cite{li2018miso} that \llg{the} above bound is \llg{asymptotically} tight at high SNR \llg{for} the single-user channel. Denote the RHS of~\eqref{eq:sum-upp} by $\mathsf{C}_{\textnormal{ap-sum-upp}}$, \llg{in the following we also present a useful result} in the paper:
\begin{IEEEeqnarray}{c}
\lim_{\sigma \rightarrow 0}\{\mathsf{C}_{\textnormal{ap-sum-upp}}+\log{\sqrt{2\pi e}\sigma}\} = h(1,\eta'),
\label{eq:uppmmm}
\end{IEEEeqnarray}
where $\eta'$ is the unique solution to
\begin{IEEEeqnarray}{c}
f(1,\eta') ={\alpha_{\text{w}}}.
\label{eq:etasolutionn}
\end{IEEEeqnarray}

\begin{remark}
The RHS of~\eqref{eq:uppmmm} corresponds to the differential entropy of the truncated exponential distribution over \llg{the} interval $[0,1]$ with parameter $\eta'$. It can be shown this term is always negative when $\eta' \in (0,\infty)$, which can be justified by the fact \llg{that} the maximal differential entropy over support $[0,1]$ is $0$ achieved by the uniform distribution over $[0,1]$.
\end{remark}

In the following, we \llg{present the} derived lower bounds on the sum-capacity. We \llg{respectively} consider the cases when $\alpha_1=\alpha_2$ and $\alpha_1\neq\alpha_2$. 
\subsubsection{On Case when $\alpha_1=\alpha_2$}
\label{subsec:alpha1=alpha2}
When $\alpha_1=\alpha_2$, it is obvious \llg{that} $\alpha_{\text{w}}=\alpha_1=\alpha_2$. Without loss of generality, we assume $\mathsf{A}_1 \leq \mathsf{A}_2$. 
\begin{theorem}
\label{prop:peakaverage}

Consider an integer $k \geq \left\lceil\frac{1}{\mathsf{A}_1}\right\rceil$. When $\alpha_1=\alpha_2$, if \llg{ $ \mathsf{E}_1 \geq \frac{1}{2k}$ and $ \mathsf{E}_2 \geq \frac{1}{2}  \left(1-\frac{\lceil \mathsf{A}_1 k \rceil}{k}\right)$,} then the sum-capacity of \llg{the} peak- and average-intensity constrained MAC is lower bounded by
\begin{IEEEeqnarray}{c}
\mathsf{C}_{\textnormal{ap-sum}} \geq \frac{1}{2}\log \left(1+\frac{l^2}{2\pi e \sigma^2}\right) ;
\label{eqLwbnd1}
\end{IEEEeqnarray}
otherwise 
\begin{IEEEeqnarray}{c}
\mathsf{C}_{\textnormal{ap-sum}} \geq \frac{1}{2}\log \left(1+\frac{e^{2\alpha^* \eta^*l}}{2\pi e \sigma^2} \left( \frac{1-e^{-\eta^*l}}{\eta^*} \right)^2 \right),
\label{eqLwbnd2}
\end{IEEEeqnarray}
where $l=\frac{1}{k}+1-\frac{\lceil \mathsf{A}_1k \rceil}{k}$, 
\begin{IEEEeqnarray}{rCl}
\eta^*= \argmin_{\eta>0} \left\{f\left(\frac{1}{k},\eta\right) \leq \mathsf{E}_1,
 g(l,\eta,k l) \leq \mathsf{E}_2   \right\},
\label{eq:valueofeta}
\end{IEEEeqnarray} 
and 
\begin{IEEEeqnarray}{rCl}
\alpha^*= f(l,\eta^*).
\end{IEEEeqnarray}
\end{theorem}

\begin{IEEEproof}
To prove~\eqref{eqLwbnd1}, let $X_1$ follow a uniform distribution over $\left[0,\frac{1}{k}\right]$, then $\mathbb{E}[X_1]=\frac{1}{2k} \leq \mathsf{E}_1$. Let $X_2$ follow a discrete uniform distribution over the support $\left\{\frac{j}{k}|j \in [k-\lceil \mathsf{A}_1 k]\right\}$. Then $\mathsf{E}[X_2] \leq \frac{k-\lceil \mathsf{A}_1 k \rceil}{2k} \leq \mathsf{E}_2$. Hence $(X_1,X_2)$ \llg{forms} an admissible \llg{input} pair, and by Proposition~\ref{prop1}, $X_1+X_2$ follows a uniform distribution over $[0,l]$. The proof is concluded by applying the EPI.

To prove~\eqref{eqLwbnd2}, let $X_1$ follow a truncated exponential distribution over $\left[0,\frac{1}{k}\right]$ with parameter $\eta^*$, then by~\eqref{eq:valueofeta} the expectation of $X_1$ satisfies
\begin{IEEEeqnarray}{rCl}
\mathbb{E}[X_1] = f\left(\frac{1}{k},\eta^*\right) \leq \mathsf{E}_1.
\end{IEEEeqnarray}
 Let $X_2$ follow a truncated geometric distribution over the support $\left\{\frac{j}{k}|j\in \left[k-\left\lceil \mathsf{A}_1  k \right\rceil\right]\right\}$ with parameter $1-e^{-\frac{\eta^*}{k}}$, then by~\eqref{eq:valueofeta} the expectation of $X_2$ satisfies
\begin{IEEEeqnarray}{rCl}
\mathbb{E}[X_2] =  g(l,\eta^*,k l) \leq \mathsf{E}_2.
\end{IEEEeqnarray}
\llg{Hence} $(X_1,X_2)$ forms an admissible input pair. By Proposition~\ref{prop9}, $X_1+X_2$ follows a truncated exponential distribution over $[0,l]$ with parameter $\eta^*$, and we have
\begin{IEEEeqnarray}{rCl}
\mathsf{h}(X_1+X_2)=h(1,\eta^*).
\end{IEEEeqnarray}
The proof is concluded by applying the EPI.
\end{IEEEproof}
\begin{remark}
\llg{Optimizing over the feasible choices of $k$ in Theorem~$3$ may get a better sum-capacity lower bound.} By the monotonicity of $f\left(\frac{1}{k},\eta\right)$ and $ g(l,\eta,k l)$ with respect to $\eta$, one of the constraints at the RHS of~\eqref{eq:valueofeta} must be tight \llg{at $\eta^*$}. For fixed $\mathsf{E}_1$ and $\mathsf{E}_2$, it can be shown \llg{that} there exists a threshold integer $k_0$  such that when $k \leq k_0$, the first constraint is tight, while when $k > k_0$, the second constraint is tight \llg{at $\eta^*$}.  
\end{remark}
The following corollary gives a more explicit bound when $\mathsf{A}_1=\frac{1}{k}$ with \llg{the} integer $k \geq 2$.
\begin{corollary}
\label{cor:lowbnd}
When $\alpha_1=\alpha_2$, and $\mathsf{A}_1=\frac{1}{k}$ with integer $k \geq 2$, the sum-capacity of \llg{the} peak- and average-intensity constrained MAC is lower bounded by
\begin{IEEEeqnarray}{c}
\mathsf{C}_{\textnormal{ap-sum}} \geq \frac{1}{2}\log \left(1+\frac{e^{2\alpha' \eta'l}}{2\pi e \sigma^2} \left( \frac{1-e^{-\eta'l}}{\eta'} \right)^2 \right),
\label{eq:lwbnd1}
\end{IEEEeqnarray}
where $l=1-\frac{1}{k}+\frac{1}{nk}$ with \llg{the} integer $n \geq \left\lceil\frac{2}{\alpha_{\textnormal{w}}} \right\rceil$, and $\eta'$ is the unique solution to 
\begin{IEEEeqnarray}{rCl}
\label{eq:constx2}
 g(l,\eta',nk  l) =\left(1-\frac{1}{k}\right)\alpha_{\textnormal{w}},
\end{IEEEeqnarray}
and 
\begin{IEEEeqnarray}{rCl}
\alpha'= f(1,\eta'l);
\end{IEEEeqnarray}
and
\begin{IEEEeqnarray}{c}
\mathsf{C}_{\textnormal{ap-sum}} \geq \frac{1}{2}\log \left(1+\frac{e^{2\alpha^* \eta^*}}{2\pi e \sigma^2} \left( \frac{1-e^{-\eta^*}}{\eta^*} \right)^2 \right),
\label{eq:lwbnd2}
\end{IEEEeqnarray}
where $\eta^*$ is the unique solution to 
\begin{IEEEeqnarray}{c}
f\left(\frac{1}{k},\eta^*\right)=\frac{\alpha_{\textnormal{w}}}{k},
\label{eq:etasolu}
\end{IEEEeqnarray}
and 
\begin{IEEEeqnarray}{rCl}
\alpha^*= f(1,\eta^*).
\end{IEEEeqnarray}
\end{corollary}
\begin{IEEEproof}
To prove~\eqref{eq:lwbnd1}, let $X_1$ \llg{follow} a truncated exponential distribution over the support $\left[0,\frac{1}{nk}\right]$ with parameter $\eta'$, and then by the fact $n \geq \left\lceil\frac{2}{\alpha_{\text{w}}} \right\rceil$,
\begin{IEEEeqnarray}{rCl}
\mathbb{E}[X_1]= f\left(\frac{1}{nk},\eta'\right)  \leq \frac{1}{2nk} \leq \frac{\alpha_{\text{w}}}{k} =\mathsf{E}_1.
\end{IEEEeqnarray}
Let $X_2$ \llg{follow} a truncated geometric distribution over the support $\left\{\frac{j}{n k}|j \in [(k-1) n] \right\}$ with parameter $1-e^{-\frac{\eta'}{k}}$, then by~\eqref{eq:constx2}
\begin{IEEEeqnarray}{rCl}
\mathbb{E}[X_2]=   g(l,\eta',nk  l) = \left(1-\frac{1}{k}\right)\alpha_{\text{w}} = \mathsf{E}_2.
\end{IEEEeqnarray}
Hence $(X_1,X_2)$ forms an admissible input pair. By Proposition~\ref{prop9}, $X_1+X_2$ follows a truncated exponential distribution over $[0,l]$ with parameter $\eta'$, \llg{and} we have 
\begin{IEEEeqnarray}{rCl}
\mathsf{h}(X_1+X_2)=h(l,\eta').
\end{IEEEeqnarray}
Then~\eqref{eq:lwbnd1} can be proved by applying the EPI.

 Now we prove~\eqref{eq:lwbnd2}. Let $X_1$ follow a truncated exponential distribution over $\left[0,\frac{1}{k}\right]$ with parameter $\eta^*$, then by~\eqref{eq:etasolu}, 
\begin{IEEEeqnarray}{rCl}
\mathbb{E}[X_1]= f\left(\frac{1}{k},\eta^*\right)=\frac{\alpha_{\text{w}}}{k} =\mathsf{E}_1.
\label{eq:eqeq1}
\end{IEEEeqnarray}
Let $X_2$ follow a truncated geometric distribution over \llg{the} support $\left\{\frac{j}{k} | j\in[k-1] \right\}$ with parameter $1-e^{-\frac{\eta^*}{k}}$. Now we verify $X_2$ satisfies the average-intensity constraint. By Proposition~\ref{prop9}, $X_1+X_2$ follows a truncated exponential distribution over $[0,1]$ with parameter $\eta^*$. Then
\begin{IEEEeqnarray}{rCl}
\mathbb{E}[X_1+X_2] = f(1,\eta^*) =k f\left(\frac{1}{k},\eta^*k\right) \leq k f\left(\frac{1}{k},\eta^*\right)= \alpha_{\text{w}}.
\end{IEEEeqnarray}
We obtain
\begin{IEEEeqnarray}{rCl}
\mathbb{E}[X_2]= \mathbb{E}[X_1+X_2] - \mathbb{E}[X_1] \leq \alpha_{\text{w}}-\frac{\alpha_{\text{w}}}{k}=\left(1-\frac{1}{k}\right)\alpha_{\text{w}}=\mathsf{E}_2.
\label{eq:eqeq2}
\end{IEEEeqnarray}
\llg{Hence} $(X_1,X_2)$ forms an admissible input pair, and 
\begin{IEEEeqnarray}{rCl}
\mathsf{h}(X_1+X_2)=h(1,\eta^*).
\end{IEEEeqnarray}
The proof is concluded by applying the EPI.
\end{IEEEproof}
Now we consider the asymptotic \llg{sum-capacity} at high SNR. Before presenting the result, we first denote the terms at the RHS of~\eqref{eqLwbnd1} or~\eqref{eqLwbnd2} by $\mathsf{C}_{\text{ap-sum-low}}$. The following proposition characterizes the asymptotic gap between $\mathsf{C}_{\text{ap-sum-low}}$ and $\mathsf{C}_{\text{ap-sum-upp}}$ in~\eqref{eq:uppmmm} at high SNR.

\begin{proposition}
\label{eq:lowboundd2}
Consider an integer $k$ satisfying $k \geq \left\lceil\frac{1}{\mathsf{A}_1}\right\rceil$, and 
\begin{IEEEeqnarray}{c}
f\left(\frac{1}{k},\frac{\eta'}{l}\right) \leq \mathsf{E}_1,
\label{eq:x1const}
\end{IEEEeqnarray}
where $\eta'$ is defined as in~\eqref{eq:etasolutionn}. Then when $\alpha_1=\alpha_2$, we have
\begin{IEEEeqnarray}{c}
\lim_{\sigma\rightarrow 0} \left\{ \mathsf{C}_{\textnormal{ap-sum-upp}} -\mathsf{C}_{\textnormal{ap-sum-low}} \right\} \leq  \log\frac{1}{l} < \log 2,
\label{eq:asympttt}
\end{IEEEeqnarray}
where $l=\frac{1}{k}+1-\frac{\lceil \mathsf{A}_1k \rceil}{k}$.
\end{proposition}
\begin{IEEEproof}
If $k  \mathsf{E}_1 \geq \frac{1}{2}$ and $k  \mathsf{E}_2 \geq \frac{1}{2}  (k-\lceil \mathsf{A}_1 k \rceil)$, then by~\eqref{eqLwbnd1}, 
\begin{IEEEeqnarray}{c}
\lim_{\sigma \rightarrow 0}\{\mathsf{C}_{\textnormal{ap-sum-low}}+\log{\sqrt{2\pi e}\sigma}\} = \log l.
\end{IEEEeqnarray}
Combined with~\eqref{eq:uppmmm}, we have
\begin{IEEEeqnarray}{rCl}
\lim_{\sigma\rightarrow 0} \left\{ \mathsf{C}_{\textnormal{ap-sum-upp}} -\mathsf{C}_{\textnormal{ap-sum-low}} \right\} &=& h(1,\eta') - \log l  \\
&<& -\log l.
\label{eq:calc}
\end{IEEEeqnarray}

Otherwise, let $X_1$ follow a truncated exponential distribution over \llg{the} support $\left[0,\frac{1}{k}\right]$ with parameter $\eta=\frac{\eta'}{l}$, then by~\eqref{eq:x1const} we have $\mathbb{E}[X_1] \leq \mathsf{E}_1$. Let $X_2$ follow a truncated geometric distribution over the support $\left\{\frac{j}{k}|j \in [1-\lceil \mathsf{A}_1 k \rceil]\right\}$. Following the same arguments as in Eq.~\eqref{eq:eqeq1}-\eqref{eq:eqeq2}, we can show 
 $\mathbb{E}[X_2] \leq \left(l-\frac{1}{k}\right) \alpha_{\text{w}} \leq \mathsf{E}_2$. Hence $(X_1,X_2)$ forms an admissible input pair, and $\eta \geq \eta^*$, where $\eta^*$ is defined as in~\eqref{eq:valueofeta}. With a slight abuse of notation, it can be shown $\mathsf{C}_{\textnormal{ap-sum-low}}$ is monotonically decreasing with $\eta^*$, then 
\begin{IEEEeqnarray}{c}
\mathsf{C}_{\textnormal{ap-sum-low}} \geq \frac{1}{2}\log \left(1+\frac{e^{2\alpha \eta l}}{2\pi e \sigma^2} \left( \frac{1-e^{-\eta l}}{\eta} \right)^2 \right).
\end{IEEEeqnarray}
By substituting $\eta=\frac{\eta'}{l}$ and some algebraic manipulations, we have
\begin{IEEEeqnarray}{c}
\lim_{\sigma \rightarrow 0}\{\mathsf{C}_{\textnormal{ap-sum-low}}+\log{\sqrt{2\pi e}\sigma}\} \geq h(1,\eta') + \log l.
\end{IEEEeqnarray}
Combined with~\eqref{eq:uppmmm}, we have
\begin{IEEEeqnarray}{c}
\lim_{\sigma\rightarrow 0} \left\{ \mathsf{C}_{\textnormal{ap-sum-upp}} -\mathsf{C}_{\textnormal{ap-sum-low}} \right\} \leq -\log l.
\label{eq:calc}
\end{IEEEeqnarray}
\end{IEEEproof}
In the following we present a lower bound based on \llg{Proposition~\ref{proplowerbndsingle}}.
\begin{proposition}
When $\alpha_1=\alpha_2$, the sum-capacity of \llg{the} peak- and average-intensity constrained MAC is lower bounded by
\begin{IEEEeqnarray}{rCl}
\mathsf{C}_{\textnormal{ap-sum}}  &\geq& \sup_{t \in (0,\eta^*)} \bigg\{ \log \frac{1-{(1-t)}^{K}}{t} - \left(K-1+\frac{1}{t}-\frac{K}{1-(1-t)^{K}}\right) \log (1-t) \nonumber \\
&&\hspace{3.2cm}- R\left(2 \left(K_1-1+\frac{1}{t}-\frac{K_1}{1-(1-t)^{K_1}}\right)  \frac{\sigma}{\mathsf{E}_1}\right)\bigg\},
\label{eq:discrepeakaverage}
\end{IEEEeqnarray}
where $K=K_1 n$ with \llg{the} integer $K_1 \geq 2$ and $n$ satisfying
\begin{IEEEeqnarray}{rCl}
n=\argmax_{m \in \mathbb{N}, \; m \leq \left\lfloor \frac{1}{\mathsf{A}_1} \right\rfloor}\left\{ \frac{1}{t}-\frac{m K_1}{1-(1-t)^{m K_1}}+m K_1-1\leq \frac{\alpha}{2l} \right\},
\label{eq:truncaexp2}
\end{IEEEeqnarray}
and $l$ being the unique solution to 
\begin{IEEEeqnarray}{rCl}
\frac{1}{t}-\frac{K_1}{1-(1-t)^{K_1}}+K_1-1=\frac{\mathsf{E}_1}{2l};
\label{eq:truncaexp1}
\end{IEEEeqnarray}
and where $\eta^*$ is the unique solution to 
\begin{IEEEeqnarray}{rCl}
\frac{1}{\eta^*}-\frac{K_1}{1-(1-\eta^*)^{K_1}}+(1-\alpha) K_1-1=0.
\end{IEEEeqnarray}
\end{proposition}
\begin{IEEEproof}
\llg{Given} $l \in \left[\frac{\mathsf{E}_1}{K_1-1},\frac{\mathsf{A}_1}{2K_1}\right]$, let $X_1$ follow \llg{a} $K_1$-point truncated geometric distribution: 
\begin{IEEEeqnarray}{rCl}
p_{X_1}(j) = \frac{\eta  (1-\eta)^{j} }{1-(1-\eta)^{K_1}}, \quad \forall j \in [K_1-1],
\end{IEEEeqnarray}
where $\eta$ is the unique solution to
\begin{IEEEeqnarray}{rCl}
\frac{1}{\eta}-\frac{K_1}{1-(1-\eta)^{K_1}}+K_1-1=\frac{\mathsf{E}_1}{2l}.
\end{IEEEeqnarray}
Then we have $\mathbb{E}[X_1]=\mathsf{E}_1$. Let $X_2$ follow \llg{an} $n$-point truncated geometric distribution:
\begin{IEEEeqnarray}{rCl}
p_{X_2} (j)= \frac{1-(1-\eta)^{K_1}}{1-(1-\eta)^{K}}  (1-\eta)^{j  K_1}, \quad \forall j \in [n-1].
\end{IEEEeqnarray}
The largest value of the mass point in the support of $X_2$ is $2l(n-1) K_1 \leq  \frac{\mathsf{A}_1}{K_1}\bigl(\frac{1}{\mathsf{A}_1}-1 \bigr) K_1 =1-\mathsf{A}_1=\mathsf{A}_2 $. Hence $X_2$ satisfies the peak-intensity constraint. By Corollary~\ref{cor:distrunexp}, $X_1+X_2$, follows an $(n K_1)$-point truncated geometric distribution over \llg{the} support $\{2lj|j\in [n K_1-1]\}$, and we have
\begin{IEEEeqnarray}{rCl}
\mathbb{E}[X_1+X_2]=2l  \left(\frac{1}{t}-\frac{n K_1}{1-(1-t)^{n K_1}}+n K_1-1\right).
\end{IEEEeqnarray}
By~\eqref{eq:truncaexp2} and~\eqref{eq:truncaexp1}, we have
\begin{IEEEeqnarray}{rCl}
 \mathbb{E}[X_2]=\mathbb{E}[X_1+X_2]-\mathbb{E}[X_1] \leq \alpha_{\text{w}}-\mathsf{E}_1=\mathsf{E}_2. 
\end{IEEEeqnarray}
Hence $(X_1,X_2)$ forms an admissible input pair. The proof is concluded by applying~\eqref{eq:peakavermmm} in Proposition~\ref{proplowerbndsingle}.
\end{IEEEproof}
\begin{remark}
When $l \in \left(0,\frac{\mathsf{E}_1}{K_1-1}\right)$, the \llg{maxentropic} input distribution over \llg{the} support $\left\{  2lj |j \in [K_1-1] \right\}$ is the uniform distribution, rather than above truncated geometric distribution. 
\end{remark}

\begin{figure}[h]
	\centering
	\resizebox{12cm}{!}{\includegraphics{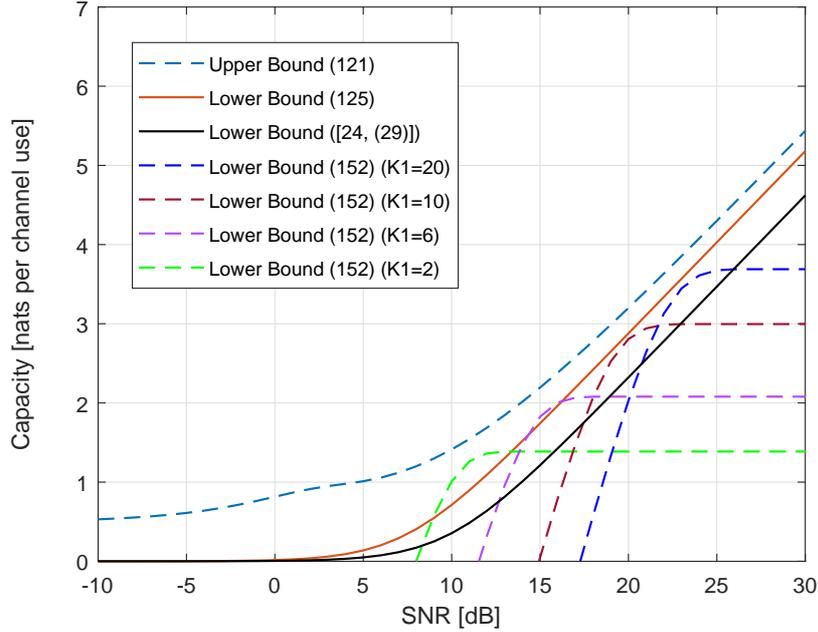}}
	\centering \caption{A two-user peak- and average-intensity constrained MAC channel ($\mathsf{A}_1=0.3$, and $\alpha_1=\alpha_2=0.4$).}
  \label{fig:5}
\end{figure}

\begin{figure}[h]
	\centering
	\resizebox{12cm}{!}{\includegraphics{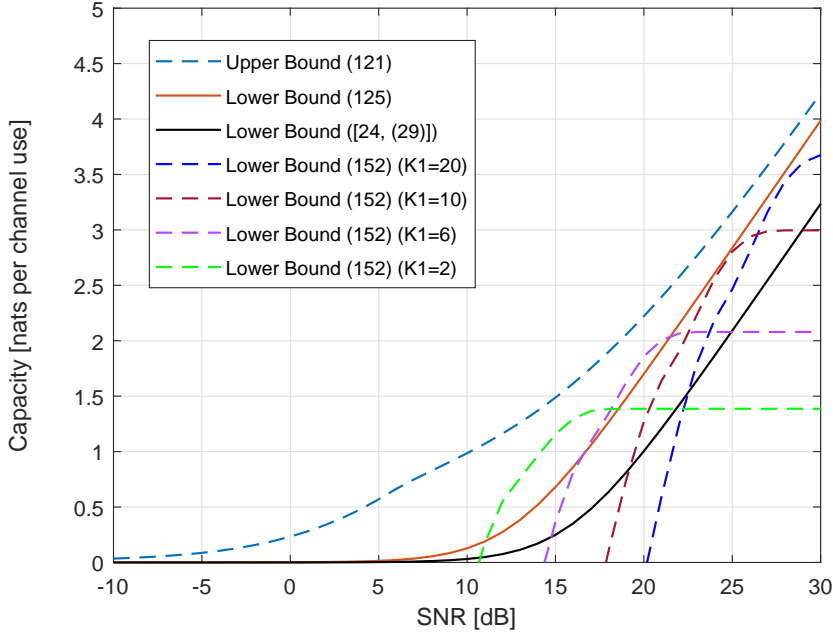}}
	\centering \caption{A two-user peak- and average-intensity constrained MAC channel ($\mathsf{A}_1=0.3$, and $\alpha_1=\alpha_2=0.1$).}
  \label{fig:52}
\end{figure}

Figure~\ref{fig:5} and~\ref{fig:52} illustrate \llg{the} above bounds by respectively letting $\alpha_1=\alpha_2=0.4$ and $\alpha_1=\alpha_2=0.1$ while fixing $\mathsf{A}_1=0.3$. At high SNR, compared with \llg{the} existing bounds in~\cite[Eq. ($69$)]{chaaban2017capacity}, the lower bound in~\eqref{eqLwbnd2} in both cases gives \llg{a} better approximation to the \llg{the} \llg{sum-capacity}, and within $0.4$ bits to the upper bound in~\eqref{eq:sum-upp}. By numerically evaluating different values of $\alpha_1=\alpha_2$, we find this lower bound is always much tighter than the given $1$-bit gap in~\eqref{eq:asympttt}. At moderate SNR, the lower bound in~\eqref{eq:discrepeakaverage} also gives a better approximation to the \llg{sum-capacity}. 

\subsubsection{Discussion on Case when $\alpha_1 \neq \alpha_2$}

When $\alpha_1 \neq \alpha_2$, without loss of generality, we assume $\alpha_1 < \alpha_2$. In the following, we first \llg{present} a result in \llg{some special cases} of $\alpha_1$ and $\alpha_2$.
\begin{proposition}
\label{prop:specialalpha12}
When $\alpha_j=\frac{\sum_{k \in \mathcal{I}_{\mathsf{A}_j}} \frac{2^{-k}} {1+e^{\eta' 2^{-k}}}}{\mathsf{A}_j},\,\, j=1,2$, the sum-capacity of \llg{the} peak- and average-intensity constrained MAC is lower bounded by
\begin{IEEEeqnarray}{c}
\mathsf{C}_{\textnormal{ap-sum}} \geq \frac{1}{2}\log \left(1+\frac{e^{2\alpha_{\textnormal{w}} \eta'}}{2\pi e \sigma^2} \left( \frac{1-e^{-\eta'}}{\eta'} \right)^2 \right),
\label{eq:symmter}
\end{IEEEeqnarray}
where $\eta'$ is the solution to 
\begin{IEEEeqnarray}{rCl}
f(1,\eta')=\alpha_{\textnormal{w}}.
\end{IEEEeqnarray}
\end{proposition}
\begin{IEEEproof}
Let  $X_1=\sum_{j \in \mathcal{I}_{\mathsf{A}_1}}B_j  2^{-j} $, and $X_2=\sum_{j \in {{\mathcal{I}_{\mathsf{A}_2}}}}B_j 2^{-j} $ with $B_j$'s being defined as in Proposition~\ref{prop:truncatedexp}. By Proposition~\ref{prop:truncated}, $(X_1,X_2)$ forms an admissible input pair, and $X_1+X_2$ follows the truncated exponential distribution over $[0,1]$ with parameter $\eta'$. The proof is concluded by applying the EPI.
\end{IEEEproof}
\llg{When $\alpha_1$ and $\alpha_2$ satisfy the conditions in Proposition~\ref{prop:specialalpha12}, it can be directly shown that as $\sigma \rightarrow 0$, \llg{the} lower bound~\eqref{eq:symmter} asymptotically matches the upper bound~\eqref{eq:sum-upp}, thus characterizing the high-SNR asymptotic sum-capacity in these cases.} 

For the general case when $\alpha_1\neq \alpha_2$, instead of directly bounding the sum-capacity, we consider the following MAC channel:
\begin{IEEEeqnarray}{rCl}
Y'=X_1'+X_2'+Z,
\label{eq:channel model2}
\end{IEEEeqnarray}
where $\mathsf{supp} X_1' \subset \left[0,\frac{\mathsf{E}_1}{\alpha'}\right]$ and $\mathbb{E}[X_1'] \leq \mathsf{E}_1$; $\mathsf{supp} X_2' \subset [0,\mathsf{A}_2]$ and $\mathbb{E}[X_2'] \leq \mathsf{A}_2  \alpha'$ with $\alpha' \in [\alpha_1,\alpha_2]$. 

It is direct to see \llg{that} for any $\alpha' \in [\alpha_1,\alpha_2]$, the sum-capacity of the channel model~\eqref{eq:channel model2} can serve as a lower bound of the original channel model~\eqref{eq:channelmodel}. Since in~\eqref{eq:channel model2}, $\alpha_1'=\alpha_2'=\alpha'$, the results in Section~\ref{subsec:alpha1=alpha2} can \llg{be directly applied here}. Then we can further refine derived capacity lower bounds by optimizing $\alpha'$ over $[\alpha_1,\alpha_2]$. 

\begin{figure}[h]
	\centering
	\resizebox{12cm}{!}{\includegraphics{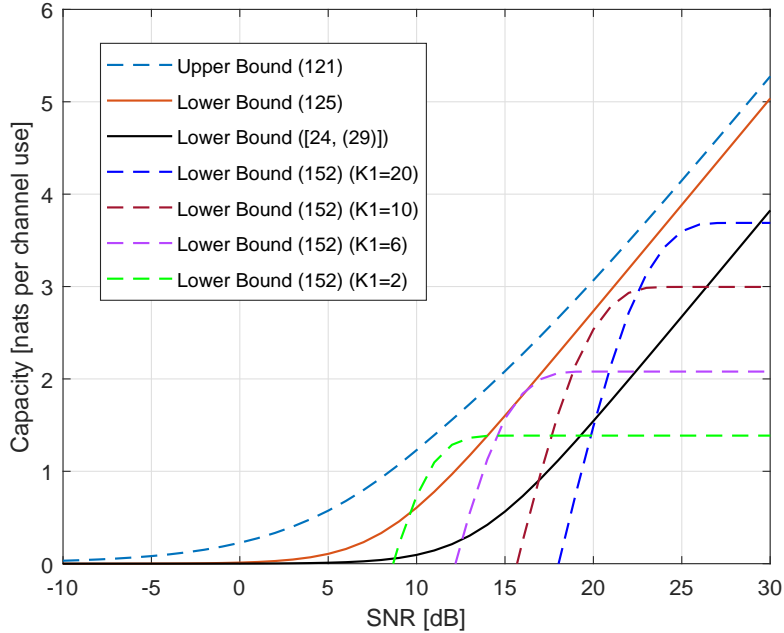}}
	\centering \caption{A two-user peak- and average-intensity constrained MAC channel ($\mathsf{A}_1=0.3$, $\alpha_1=0.1$, and $\alpha_2=0.4$).}
  \label{fig:7}
\end{figure}

Although for general $\alpha_1$ and $\alpha_2$, we cannot \llg{characterize the high-SNR asymptotic sum-capacity}, by numerically evaluating the above proposed method, we find the derived lower bounds still give good approximations to the \llg{the sum-capacity} at moderate and high SNR. Figure~\ref{fig:7} shows an example when $\alpha_1=0.1$, $\alpha_2=0.4$, and $\mathsf{A}_1=0.3$. At high SNR, the lower bound~\eqref{eqLwbnd2} is fairly close to the upper bound~\eqref{eq:sum-upp}. At moderate SNR, the lower bound~\eqref{eq:discrepeakaverage} gives a better approximation to the sum-capacity compared with \llg{the} existing sum-capacity bounds. 

\section{Conclusion and Future Works} 
\label{sec:conclusion}
In this paper, we propose methods \llg{for} decomposing a continuous or discrete random variable as a sum of two independent random variables. The distribution of the considered random variable achieves the maximum entropy under certain support or moment constraints. With the decomposition results, we derive several capacity lower bounds on the sum-capacity of two-user MAC when the input is subject to the peak- or/and average-intensity constraints. Compared with \llg{the} existing results, at moderate SNR, some bounds give better approximations to the sum-capacity. At high SNR, some bounds are asymptotically tight or fairly close to the sum-capacity, thus determining or improving \llg{evaluations} on \llg{the} high-SNR asymptotic sum-capacity.

 \llg{In the peak- and average-intensity constrained case}, \llg{in order to apply the decomposition results, some lower bounds are derived by letting channel inputs satisfy slightly tighter constraints than the given ones, which may lose the optimality at high SNR. Evaluations on sum-capacity may be improved by finding admissible and decomposable distributions to make the inputs satisfy exactly the given constraints.}
Furthermore, it is worth mentioning that the idea of decomposing a certain random variable (or vector) as a sum of two or several dependent or independent random \llg{variables} (or \llg{vectors}) under certain support or moment constraints \llg{has also implicitly} applied to \llg{some single-user} wireless optical intensity channels~\cite{limoserwangwigger20_1,
moserwangwigger18_3,chen2021MISO}. A natural extension of the topic is to apply the idea to other optical intensity networks, such as broadcast channels, interference channels, and relay channels. We leave this for future study.
\appendix
\section{Proof of Lemma~$1$}
\label{app:lem11}
\llg{
We only need to show that $\forall y \in (0,l)$, $y \in \mathcal{A}$. When $y \in (0,l)$, denote $j_1$ as the smallest index satisfying $a_{j_1} \leq y$. Denote $j_2$ as the smallest index satisfying $j_2>j_1$ and $a_{j_1}+a_{j_2}\leq y$ if $j_2$ exists. Repeat this procedure, i.e., assuming $j_1,j_2,\ldots,j_m$ is chosen, then we choose $j_{m+1}$ to be the smallest index satisfing $j_{m+1}>j_m$ and $\sum_{k=1}^{m}a_{j_k}+a_{j_{m+1}}\leq y$ if $j_{m+1}$ exists. By this process, we get an index sequence $j_1,j_2,\cdots$. 

If the index sequence is finite, assuming the sequence length is $n$, then by the process, we have $\sum_{k=1}^{n}a_{j_k} \leq y$. Also, since the process stops at $j_n$, then for any integer $m>j_n$, we have $a_m+ \sum_{k=1}^{n}a_{j_k}>y$. By the fact $\lim_{m\rightarrow \infty}a_m =0$, we get $\sum_{k=1}^{n}a_{j_k} = y$. Hence $y \in \mathcal{A}$.

If the index sequence is infinite, since $0<y<l$, there must be integers that do not appear in the sequence $j_1,j_2,\cdots$. We denote these integers by $\{q_1,q_2,\ldots\}$. Now we first prove that the set $\{q_1,q_2,\ldots\}$ contains an infinite number of integers. Assume the set is finite, and denote $q$ as the largest integer in it. Since $q \notin \{j_1,j_2,\cdots\}$, we have 
\begin{IEEEeqnarray}{rCl}
a_q+ \sum_{j_k < q}a_{j_k}>y.
\label{eq:cont1} 
\end{IEEEeqnarray}
Also, by the process we have 
\begin{IEEEeqnarray}{rCl}
\sum_{k=1}^{\infty}a_{j_k} \leq y,
\end{IEEEeqnarray}
which is equivalent to 
\begin{IEEEeqnarray}{rCl}
\sum_{j_k < q}a_{j_k}+ \sum_{k=q+1}^{\infty}a_{k}  \leq y.
\label{eq:cont2}
\end{IEEEeqnarray}
Combining~\eqref{eq:cont1} with~\eqref{eq:cont2}, we get $a_q > \sum_{k=q+1}^{\infty}a_{k}=r_q$. Contradiction occurs. Hence the set $\{q_1,q_2,\cdots\}$ must be an infinite set. Then for any $q_t \in \{q_1,q_2,\cdots\}$, we have 
\begin{IEEEeqnarray}{rCl}
a_{q_t}+ \sum_{j_k < q_t}a_{j_k}>y.
\label{eq:cont3}
\end{IEEEeqnarray}
Also, by the process we have 
\begin{IEEEeqnarray}{rCl}
\sum_{j_k < q_t}a_{j_k} \leq y.
\label{eq:cont4}
\end{IEEEeqnarray}
Combing~\eqref{eq:cont3} with~\eqref{eq:cont4}, and by the fact $\lim_{t\rightarrow \infty}a_{q_t} =0$, we get $\sum_{k=1}^{\infty}a_{j_k} = y$. Hence $y \in \mathcal{A}$. The proof is concluded.
}

\section{Proof of~\llg{Lemma~$2$}}
\label{app:1a}
The c.f. of $2^j B_j$ is
\begin{IEEEeqnarray}{c}
\phi_{2^j  B_j}(t) = \mathsf{E}[e^{it2^j  B_j}]= \frac{1+e^{-(\lambda-it)2^j}}{1+e^{-\lambda 2^j}}.
\end{IEEEeqnarray}
Then the c.f. of $U$ is 
\begin{IEEEeqnarray}{c}
\label{eq:cfu}
\phi_{U}(t) = \prod_{j=-\infty}^{\infty} \frac{1+e^{-(\lambda-it)2^j}}{1+e^{-\lambda 2^j}}.
\end{IEEEeqnarray} 
Using the relation
\begin{IEEEeqnarray}{c}
\prod_{j=0}^{n}(1+e^{z2^j})=  \frac{1-e^{z2^{n+1}}}{1-e^z},
\end{IEEEeqnarray} 
we have 
\begin{IEEEeqnarray}{c}
\label{eq:part1}
\prod_{j=0}^{n} \frac{1+e^{-(\lambda-it)2^j}}{1+e^{-\lambda 2^j}} = \frac{1-e^{-\lambda}}{1-e^{-(\lambda-it)}} \frac{1-e^{-(\lambda-it)2^{n+1}}}{1-e^{-\lambda2^{n+1}}} .\nonumber \\
\end{IEEEeqnarray}
Following \llg{the} above similar arguments, we can derive
\begin{IEEEeqnarray}{c}
\label{eq:part2}
\prod_{j=-1}^{-n} \frac{1+e^{-(\lambda-it)2^j}}{1+e^{-\lambda2^j}} = \frac{1-e^{-(\lambda-it)}}{1-e^{-\lambda}} \frac{1-e^{-\lambda2^{-n}}}{1-e^{-(\lambda-it)2^{-n}}} .\nonumber\\
\end{IEEEeqnarray}
\llg{By plugging~\eqref{eq:part1} and~\eqref{eq:part2} into~\eqref{eq:cfu} and letting $n\rightarrow\infty$,} we have 
\begin{IEEEeqnarray}{c}
\phi_{U}(t) = \frac{\lambda}{\lambda-it},
\end{IEEEeqnarray}
which is the c.f. of the exponential distribution with parameter $\lambda$. The proof is concluded.
\section{Proof of Eq.~\eqref{eq:lessrn}}
We first prove the sequence $(b_n)_{n \geq 1}$ is monotonically decreasing. Using the fact that \llg{the} function $f(x)=\frac{x}{1+e^x}$ is monotonically increasing on $[0,1]$, and decreasing on $[2,+\infty]$, then $\forall n \in \mathbb{N}$, $a_{-n+1}>a_{-n}$, and $a_{n-1}>a_n$. Hence when $n\geq 5$, 
\begin{IEEEeqnarray}{c}
b_n =a_{-n+2}+a_{n-2} > b_{n+1}=a_{-n+1}+a_{n+1}.
\end{IEEEeqnarray}  
When $n\leq 4$, we can numerically verify 
\begin{IEEEeqnarray}{c}
b_n>b_{n+1}.
\end{IEEEeqnarray}
Now we prove eq.~\eqref{eq:lessrn}. 
When $n\geq 7$,
\begin{IEEEeqnarray}{rCl}
r_n &=& \sum_{i=n+1}^{\infty} \left(\frac{2^{-(i-2)}}{1+e^{2^{-(i-2)}}} + \frac{2^{i-2}}{1+e^{2^{i-2}}}\right) \\
&>&  \sum_{i=n+1}^{\infty} \frac{2^{-(i-2)}}{1+e^{2^{-(i-2)}}} \\
&=&  \sum_{i=n+1}^{\infty} \frac{2^{-(i-2)}}{1+e^{2^{-(i-3)}}} \frac{1+e^{2^{-(i-3)}}}{1+e^{2^{-{\llg{(i-2)}}}}} \\
&=&  \sum_{i=n+1}^{\infty} \frac{2^{-(i-2)}}{1+e^{2^{-(i-3)}}}+\sum_{i=n+1}^{\infty} \frac{2^{-(i-2)} { (e^{2^{-(i-2)}}-1)}}{1+e^{2^{-(i-3)}}} \times \nonumber\\
&&\hspace{5.5cm}  \frac{e^{2^{-(i-2)}}}{1+e^{2^{-(i-2)}}}  \nonumber \\
\\
&>&  \frac{\sum_{i=n+1}^{\infty}2^{-(i-2)}}{1+e^{2^{-(n-2)}}} + \frac{1}{6}  \sum_{i=n+1}^{\infty}2^{-2(i-2)} \label{eq:11} \\
&=& a_{-n+2} +  \frac{1}{18} 2^{-2(n-2)} \\
&>& a_{-n+2} + a_{n-2} \label{eq:22} \\
&=&b_n,
\end{IEEEeqnarray}
where \eqref{eq:11} follows from the fact $e^x-1>x$, and $\frac{x}{1+x}>\frac{1}{2}$, $\forall x>1$, and~\eqref{eq:22} from the following
\begin{IEEEeqnarray}{rCl}
a_{n-2} &=&\frac{2^{n-2}}{1+e^{2^{n-2}}} \\
  &<&\frac{  6!}{2^{6(n-2)}}2^{n-2} \\
 &=&\frac{6!}{2^{3(n-2)}} 2^{-2(n-2)}\\
&<&\frac{1}{18}  2^{-2(n-2)},\,\, \,\forall n \geq 7.
\end{IEEEeqnarray}
 When $n \leq 6$, we can numerically verify 
\begin{IEEEeqnarray}{c}
b_n <b_{n+1}+b_{n+2}<\sum_{i=n+1}^{\infty}b_i=r_n.
\end{IEEEeqnarray} 
The proof is concluded.
\section{Proof of Proposition~\ref{proplowerbndsingle}}
\label{appproplowerbndsingle}
In the \llg{average-intensity} constrained case, $\mathsf{H}(U_D)$ in~\eqref{eq:lowbndud} is maximized by letting $U_D$ follow a uniform distribution over $(K+1)$ equi-spaced points. Since $R(u)$ is a decreasing function over $u$, the maximal feasible $l=\frac{1}{2K}$. Hence~\eqref{eq:peaklbnd} is proved. 

\llg{In the average-intensity} constrained case, denote $p_j=p_{U_D}(j)$, and to satisfy the \llg{average-intensity} constraint, we have
\begin{IEEEeqnarray}{c}
\subnumberinglabel{eq:constraints}
p_j \geq 0;\\
\sum_{j=0}^{K}p_j = 1;\\
2l  \sum_{j=0}^{K}p_j j \leq \alpha.
\label{eq:avgcosnt}
\end{IEEEeqnarray}

When $l \in (0,\frac{\alpha}{K}]$, \llg{we let $U_D$ follow a uniform distribution over $(K+1)$ \llg{points}, and it is direct to verify $U_D$ satisfies the average intensity constraint~\eqref{eq:avgcosnt}. Then we have $\mathsf{H}(U_D)=\log (K+1)$.} 

When $l \in (\frac{\alpha}{K},\infty)$, by the \llg{maximum-entropy} argument, $U_D$ follows a truncated geometric distribution~\cite[Prop. $4.1$]{faridhranilovic09_1},
\begin{IEEEeqnarray}{rCl}
p_j^*= \frac{\zeta (1-\zeta)^{j-1}}{1-(1-\zeta)^{K+1}},
\end{IEEEeqnarray}
where $\zeta$ is the unique solution to 
\begin{IEEEeqnarray}{rCl}
K+\frac{1}{\zeta}-\frac{K+1}{1-{(1-\zeta)}^{K+1}} = \frac{\alpha}{2l}.
\label{eq:zetasol}
\end{IEEEeqnarray}
Hence we have
\begin{IEEEeqnarray}{rCl}
\mathsf{H}(U_D)=\log \frac{1-{(1-\zeta)}^{K+1}}{\zeta} - \frac{\alpha}{2l} \log (1-\zeta). \label{eq:limitmmm}
\end{IEEEeqnarray}

Since $R(u)$ is an increasing function over $u$, while $\mathsf{H}(U_D)$ is fixed when $l \in (0,\frac{\alpha}{K}]$, $\mathsf{I}(U_D;Y)$ is maximized when $l = \frac{\alpha}{K}$, which is also equal to the limit of the RHS of~\eqref{eq:limitmmm}. Hence we only need to consider the case when $l \in (\frac{\alpha}{K},\infty)$. Then 
\begin{IEEEeqnarray}{rCl}
\mathsf{I}(U_D;Y) &\geq& \sup_{l \in (\frac{\alpha}{K},\infty)}  \log \frac{1-{(1-\zeta)}^{K+1}}{\zeta} - \frac{\alpha}{2l} \log (1-\zeta) - R\left(\frac{\sigma}{l}\right) \label{eq:original} \\
&=& \sup_{t \in (0,1)} \log \frac{1-{(1-t)}^{K+1}}{t} - \left(K+\frac{1}{t}-\frac{K+1}{1-(1-t)^{K+1}}\right) \log (1-t) \nonumber \\
&&\hspace{2.9cm}- R\left(2 \left(K+\frac{1}{t}-\frac{K+1}{1-(1-t)^{K+1}}\right)  \frac{\sigma}{\alpha}\right); \label{eq:changevar}
\end{IEEEeqnarray}
where~\eqref{eq:changevar} follows by letting $t= \zeta$ and substituting it into~\eqref{eq:limitmmm} and~\eqref{eq:original}.

Now we consider a special case when $K=\infty$. The average-intensity constraint in~\eqref{eq:avgcosnt} is always active. By the \llg{maximum-entropy} argument again, $U_D$ follows a geometric distribution with parameter $\frac{2l}{2l+\alpha}$, and 
\begin{IEEEeqnarray}{rCl}
\mathsf{H}(U_D)= \log\left(1+\frac{\alpha}{2l}\right)+\frac{\alpha}{2l}\log\left(1+\frac{2l}{\alpha}\right).
\end{IEEEeqnarray}
Then~\eqref{eq:maxgeomtric} is proved by optimizing over \llg{the feasible choices} of $l$.

To prove~\eqref{eq:peakavermmm}, we can follow the similar arguments as in the proof of~\eqref{eq:changevar}. The only difference here is \llg{the maximal feasible value of $l$ is $\frac{1}{2K}$. When $l=\frac{1}{2K}$, let $\eta = \zeta$, and substitute it into~\eqref{eq:zetasol}, we obtain} 
\begin{IEEEeqnarray}{rCl}
\frac{1}{\eta}-\frac{K+1}{1-(1-\eta)^{K+1}}+(1-\alpha) K=0.
\end{IEEEeqnarray}
The proof is concluded.

\bibliographystyle{IEEEtran}
\bibliography{./defshort1,./biblio1}

%
%
%
%
%




\end{document}